\documentclass{ws-ijmpa}
\usepackage[super,compress]{cite}
\usepackage{graphicx,dsfont}
\begin{document}
\markboth{E. Gr\"unwald, Y. Delgado Mercado, C. Gattringer}
{Taylor- and fugacity expansion for the
effective $\mathds{Z}_3$ spin model of QCD at finite density}

%
\catchline{}{}{}{}{}
%

\title{Taylor- and fugacity expansion for the \\
effective $\mathds{Z}_3$ spin model of QCD at finite density
}

\author{Eva Gr\"unwald}
\author{Ydalia Delgado Mercado}
\author{Christof Gattringer}

\address{\vskip5mm
Institute for Physics, University of Graz, \\
Universit\"atsplatz 5, 8010 Graz, Austria 
\vskip5mm}

\maketitle

\begin{history}
\received{Day Month Year}
\revised{Day Month Year}
\end{history}

\begin{abstract}
Different series expansions in the chemical potential $\mu$ are studied and compared for an 
effective theory of QCD which has a flux representation where the complex action 
is overcome. In particular we consider fugacity series, Taylor expansion and a 
modified Taylor expansion and compare the outcome of these series
to the reference results from a Monte Carlo simulation in the flux representation 
where arbitrary $\mu$ is accessible. It is shown that for most parameter values the fugacity 
expansion gives the best approximation to the data from the flux simulation, followed by our newly proposed 
modified Taylor expansion. For the conventional Taylor expansion we find that the results coincide with the 
flux data only for very small $\mu$.
 \keywords{lattice field theory; finite density; series expansion.}
\end{abstract}

\ccode{PACS numbers: 12.38.Aw, 11.15.Ha}

\section{Introduction}
Since the initial formulation three decades ago lattice QCD has developed into a reliable quantitative tool 
for studying many low energy phenomena in QCD. However, one important issue where the lattice approach 
has essentially failed so far is its application to QCD at finite density. The reason is that at finite chemical potential 
$\mu$ the fermion determinant becomes complex and cannot be used as a probability weight in a Monte Carlo
simulation. This is know as the ''complex action problem'' or ''sign problem''.

One of the attempts to overcome the complex action problem is based on Taylor expansion in $\mu$ around the 
$\mu = 0$ theory such that the expansion coefficients can be computed with conventional Monte Carlo simulations. 
The Taylor expansion suffers from two problems: The simulations performed at $\mu = 0$ for the determination 
of the expansion coefficients may be governed by physics that is rather different from the finite density phenomena
one wants to study -- this is known as the ''overlap problem''. The second problem is the unknown and probably 
rather small region of convergence of the expansion. 

However, Taylor expansion is not the only expansion that might be used for extrapolating from $\mu = 0$ to finite $\mu$,
and other series might have better convergence and/or better overlap properties. In this paper we study the conventional 
Taylor expansion, but also the fugacity series and in addition propose a modified Taylor expansion which captures some
features of the fugacity expansion but at a much lower numerical cost. 

The three different series expansions are implemented in an effective theory of QCD containing the leading center 
symmetric and center symmetry breaking terms for Polyakov loops represented by $\mathds{Z}_3$-valued spins
on a 3-dimensional lattice. The model is often referred to as ''$\mathds{Z}_3$  spin model'' and has the advantage  
that it can be rewritten exactly to new variables, a so-called flux representation, where the partition sum has 
only real and positive terms. In the flux representation the complex action problem thus is solved and Monte Carlo
simulations are possible at arbitrary $\mu$. The results of the simulation in the flux representation then serve as 
reference data for the series expansions we want to study. 

The paper is organized as follows: In the next section we briefly review the $\mathds{Z}_3$  spin model and aspects 
of its simulation with the flux representation. Subsequently we present and work out the three series expansions 
for the spin model. Section 4 contains the numerical results and the comparison of the series expansions. 
We conclude with a discussion.
 
\section{The $\mathds{Z}_3$ spin model}

An effective theory \cite{ploopmodel_a,ploopmodel_b} for the Polyakov loop in pure gauge theory can be computed in strong coupling
and gives rise to a nearest neighbor interaction of traced SU(3) spins on a 3-dimensional lattice. Subsequently the 
Polyakov loop spins may be reduced to the center group of SU(3), i.e., to spins in $\mathds{Z}_3$.  

The fermion determinant breaks the center symmetry explicitly. The corresponding 
leading center symmetry breaking terms may be obtained from hopping expansion
which represents the fermion determinant as a set of closed loops. The shortest loop that breaks 
center symmetry is the Polyakov loop. It is also the leading contribution that couples to the chemical 
potential which gives a  different weight to forward and backward running loops.  
 
The action of the resulting  $\mathds{Z}_3$ spin model \cite{centermodel1_a,centermodel1_b,centermodel1_c,centermodel2_a,centermodel2_b} reads
\begin{equation}
S_\mu[P] \;  = \; - \sum_x \left( \tau  \sum_{\nu = 1}^{3} \left[ P_x P_{x + \hat{\nu}}^*  + c.c. \right] + \eta P_x + \bar{\eta}P_x^* \right) ,
\label{Z3effectiveaction}
\end{equation}
where the Polyakov loops are represented by the spins $P_x$ which are elements of 
$\mathds{Z}_3 = \left\lbrace 1, e^{\pm 2 i \pi /3} \right\rbrace$. 
The first sum runs over all sites $x$ of a $N^3$ lattice with periodic boundary conditions and 
$\hat{\nu}$ denotes the unit vector in $\nu$-direction. The chemical potential $\mu$ enters through $\eta = \kappa e^\mu $, 
$\bar{\eta} = \kappa e^{ -\mu}$. We remark that actually the leading $\mu$-dependent terms form the loop expansion of the
fermion determinant would come
with factors $e^{\pm \mu \beta}$, while in our effective action we have the factors $e^{\pm \mu}$. In other words, for
the spin model the chemical potential is rescaled with the inverse temperature  $\beta$.
The parameter $\tau$ in front of the nearest neighbor term is increasing with temperature, whereas 
$\kappa$ is increasing with decreasing QCD quark mass and is proportional to the number of flavors. 

The grand canonical partition sum $Z(\mu)$ is obtained as a sum over all configurations $\{ P \}$ of the 
variables, i.e., $Z(\mu) = \sum_{\{ P \}} e^{-S_\mu[P]}$.
Simple observables one may study are the expectation value of the Polyakov loop  
$\langle P_x \rangle = V^{-1} \langle \sum_x P_x \rangle  = V^{-1} \partial \ln Z / \partial \eta $ 
of the Polyakov loop and the corresponding Polyakov loop susceptibility $\chi_P$. Observables related to the 
particle number density are obtained as derivatives of $\ln Z$ with respect to the chemical potential $\mu$. They turn out 
to be related to linear combinations of $\langle P_x \rangle$ and $\chi_P$.

\begin{figure}[t]
\centering
\includegraphics[height=43mm,clip]{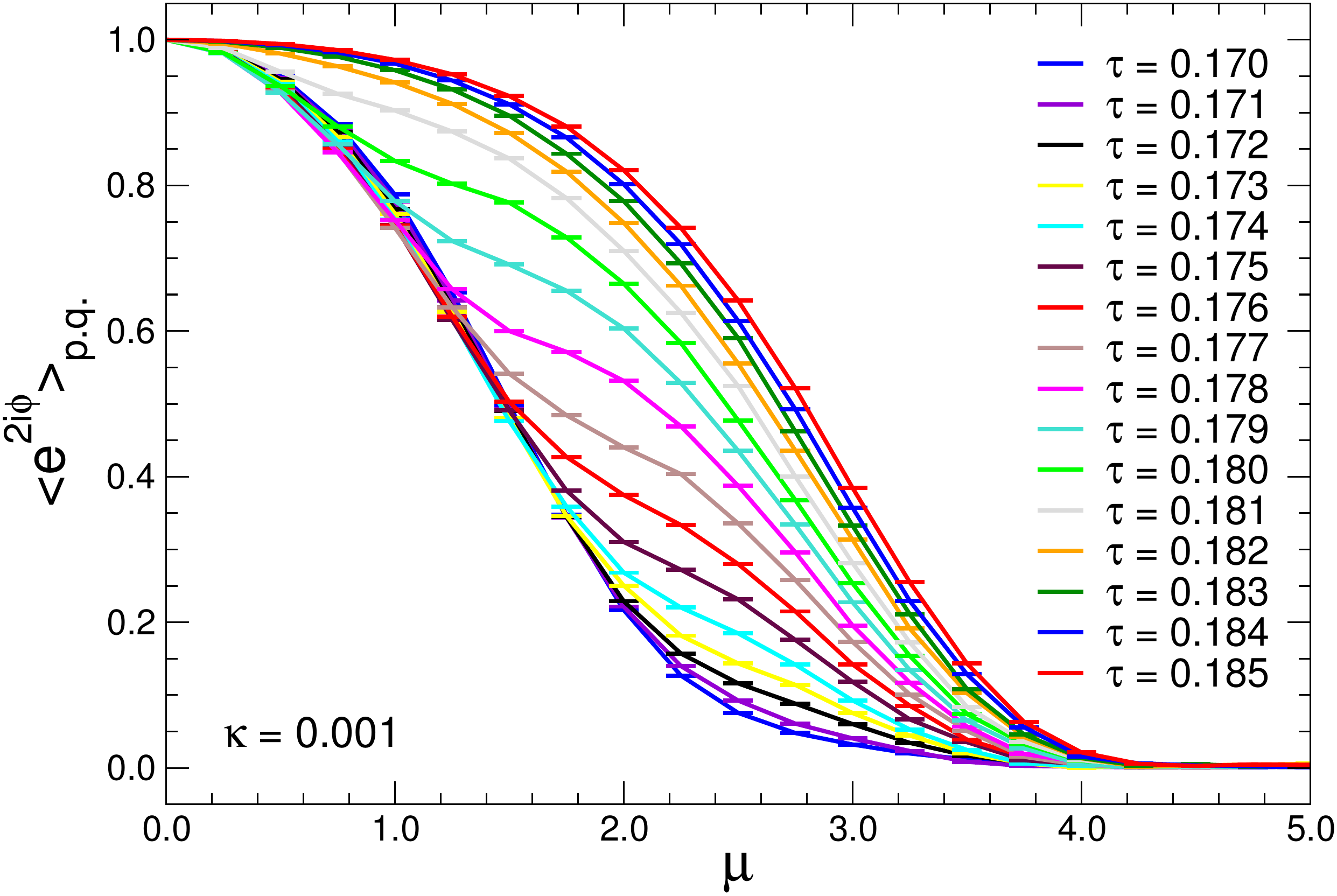}
\label{fig:figure22}
\hspace{2mm}
\centering
\includegraphics[height=43mm]{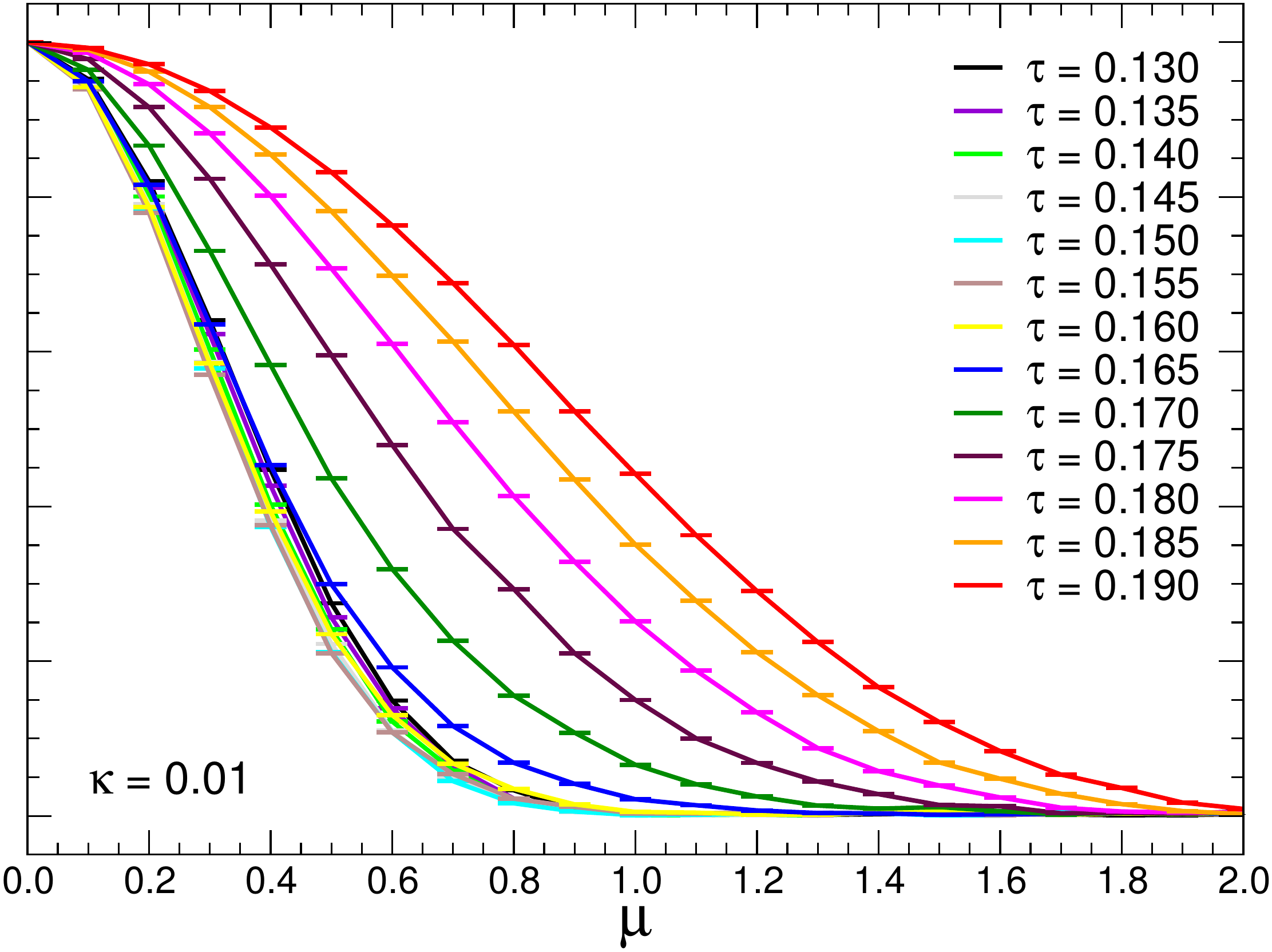}
\label{fig:figure23}
\caption{$\left\langle e^{i2 \phi} \right\rangle _ {p.q.}$ versus $\mu$ in the 
phase quenched theory for $\kappa = 0.001$ (lhs.\ plot) and 
$\kappa = 0.01$ (rhs.) on $16^3$ lattices for different values of $\tau$. 
The smaller $\tau$, the further left is the corresponding curve. Note the different 
scale on the horizontal axes of the two plots.}
\label{CAP}
\end{figure}

It is obvious that in the standard representation the action (\ref{Z3effectiveaction}) is complex for $\mu \neq 0$ and conventional 
Monte Carlo techniques fail. The severity of the complex action problem depends on the parameters  $\kappa, \tau$ and $\mu$.
Later we will see that also convergence properties of the 
various series expansions are related to the severity of the complex action problem. 
In order to assess the severity of the complex action problem, in Fig.~\ref{CAP} we show 
results for $\langle e^{i2\, \phi} \rangle _ {p.q.}$ as a function of  $\mu$ for $\kappa = 0.001$ 
(lhs.\ plot) and $\kappa = 0.01$ (rhs.) on $16^3$ lattices for different values of $\tau$.
Here we write the Boltzmann factor as $e^{-S_\mu[P]} = | e^{-S_\mu[P]} | \, e^{i \phi}$, and $\langle .. \rangle_{p.q.}$ denotes
the phase quenched expectation value, i.e., the expectation value computed with the weight $| e^{-S_\mu[P]} |$. 

From Fig.~\ref{CAP} it is obvious, that for small temperature parameter $\tau$ the complex action problem is
more severe, i.e., for smaller $\tau$ the expectation value $\langle e^{i2\, \phi} \rangle _ {p.q.}$ drops faster as a 
function of $\mu$ than it does for larger $\tau$. Furthermore, the complex action problem is more severe for larger values 
of $\kappa$, which corresponds to small quark mass. Both these findings are qualitatively the same as one expects for QCD.

The $\mathds{Z}_3$ spin model can be mapped exactly to a flux representation where the dynamical degrees of freedom are 
fluxes on the links of the lattice subject to constraints: the conservation 
of flux at the sites of the lattice, with $\kappa e^{\pm \mu}$ acting 
as sources and sinks of flux. The system can be simulated with a generalization of the Prokof'ev-Svistunov worm algorithm 
\cite{worm} and observables can be studied for arbitrary chemical potential \cite{centermodel2_a,centermodel2_b}. 
For $\mu = 0$ and sufficiently small $\kappa$ the model has a first order phase transition as a function of $\tau$ which
continues as a short first order line ending in a second order point when increasing $\mu$
 \cite{potts_a,potts_b,potts_c,Alford,Kim,centermodel2_a,centermodel2_b}. When $\kappa$ and $\mu$ are sufficiently large
one only finds a smooth crossover between $\langle P_x \rangle \sim 0$ (confinement) and $\langle P_x \rangle > 0$ 
(deconfinement). 

Having at hand a model that shows a similar complex action problem as QCD and at 
the same time has a flux representation where simulations at finite $\mu$ 
are possible without problems, is perfect for assessing the 
applicability of series expansions in the chemical potential. We will see that all three series, fugacity, Taylor and improved 
Taylor expansion are straightforward to implement in the $\mathds{Z}_3$ spin model and the flux simulation provides the 
reference data to assess the results from the series expansions.

\section{Three types of series expansions}

In this section we discuss the three series expansions we study in this paper. First we present the series expansion in general terms
such that the formulation for QCD is transparent and then work out the specific form for the $\mathds{Z}_3$ spin model. 

\subsection{Fugacity expansion}

The fugacity expansion expresses the grand canonical partition sum $Z(\mu)$ as a Laurent series in the fugacity parameter 
$e^{\mu \beta}$,  
\begin{equation}
Z(\mu) \;  = \; \sum_{q \in \mathds{Z}} e^{\mu \beta q} \, Z_q \;\;\; , \; \; \; \; 
Z_q \; = \;  \int_{-\pi}^\pi \! \frac{d \varphi}{2\pi} \; e^{- i q \varphi} \; Z(\mu = i \varphi / \beta) \; ,
\label{Z_fug}
\end{equation}
where the sum runs over all net particle numbers $q$, which give the difference between particles and anti-particle numbers. 
The expansion coefficients are the canonical partition sums $Z_q$ that may be obtained as the Fourier moments of the grand 
canonical partition sum with imaginary chemical potential $\mu = i \varphi / \beta$. In full QCD the grand canonical partition 
function is a path integral over all gauge configurations, $Z(\mu) = \int D[U] \, e^{-S_g[U]} \; \mbox{det} \, D[U,\mu]$, where 
$S_g[U]$ is the gauge action and $\mbox{det} \, D[U,\mu]$ the (grand canonical) fermion determinant.  
Since only the fermion determinant depends on the chemical potential, one can directly expand the fermion determinant,
\begin{equation}
\mbox{det} \, D[U,\mu] \; = \; \sum_{q \in \mathds{Z}} e^{\mu \beta q} \, D_q[U]\;\;\; , \; \; \; \; 
D_q[U] \; = \;  \int_{-\pi}^\pi \! \frac{d \varphi}{2\pi} \; e^{- i q \varphi} \; \mbox{det} \, D[U, \mu = i \varphi / \beta] \; ,
\label{candet}
\end{equation}
where the $D_q[U]$ are the so-called canonical determinants, i.e., the usual grand canonical determinants projected to a 
fixed net quark number $q$. The canonical determinants $D_q$ are the Fourier moments of the fermion determinant with respect 
to imaginary chemical potential $\mu = i \varphi / \beta$. The corresponding Fourier integrals need to be evaluated numerically, 
which is a very expensive calculation, despite the fact that 
a dimensional reduction formula \cite{fugacity1} may be used to speed up the 
evaluation of the $D_q$ \cite{fugacity1,fugacity2_a,fugacity2_b}.  Thus it seems reasonable to first test the fugacity expansion in a model 
where reliable results at finite $\mu$ are available, which is the issue of this paper.

For the $\mathds{Z}_3$ spin model we can implement the fugacity expansion in exactly the same way as in QCD
(with the difference that in the spin model the inverse temperature $\beta$ is absorbed in the definition of $\mu$). 
The grand canonical partition sum is
\begin{equation}
Z(\mu) \; = \; \sum_{\{P\}} e^{ \,  \tau \sum_{x,\nu} \left[ P_x P_{x + \hat{\nu}}^*  + c.c. \right]} \;
e^{ \,  \kappa e^\mu \,  M  \, + \, \kappa e^{-\mu} M^* } \; ,
\end{equation}
where we introduced the magnetization
\begin{equation}
M \; = \; \sum_x P_x \; \equiv \; R \, e^{i\theta} \; ,
\end{equation}
which for later use is also written in polar form. The fugacity expansion (\ref{Z_fug}) has the same 
form and we can write the
canonical partition sums $Z_q$ in the form
\begin{equation}
Z_q \; = \;  \sum_{\{P\}} e^{ \,  \tau \sum_{x,\nu} \left[ P_x P_{x + \hat{\nu}}^*  \, + \; c.c. \right]} \; D_q \; ,
\end{equation}
with 
\begin{equation}
D_{q} \; = \;    \int_{-\pi}^\pi \! \frac{d \varphi}{2\pi} \;   e^{-i\varphi \, q} \; 
\exp \left( \kappa e^{i\varphi} M + \kappa e^{-i\varphi} M^* \right)  \; = \; 
 e^{i \theta q} \,  I_q \left( 2 \kappa R \right) ,
\end{equation}
where in the second step we used $M = R e^{i\theta}$ and evaluated the $\varphi$-integral, giving rise 
to the modified Bessel functions $I_q$. Obviously the $D_q$ are the analogues of the canonical determinants 
of QCD, i.e., the fermion determinant projected to a fixed net quark number sector. 
 
It is obvious that the $D_q$ must decrease with increasing $q$, such that the fugacity series (\ref{Z_fug}) converges. 
In a practical implementation the fugacity series must be truncated to values $q$ of the particle number in some interval 
with a lower and an upper bound, i.e., $q_l \leq q \leq q_u$. The analysis of the size distribution of the $D_q$ is necessary 
for obtaining a reasonable estimate for $q_l$ and $q_u$. 

\begin{figure}[t]
\includegraphics[height=42mm]{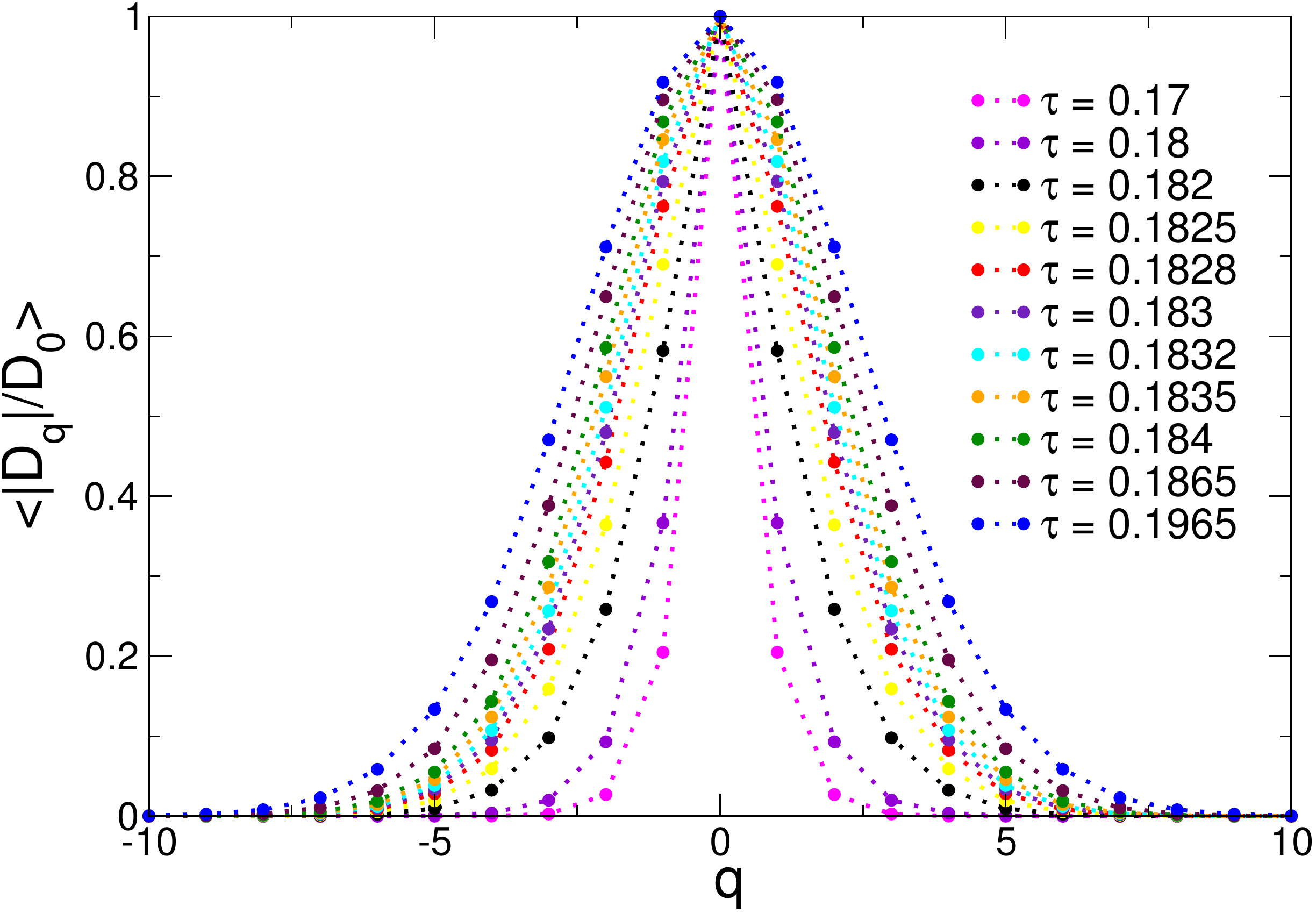}
\hspace{2mm}
\includegraphics[height=42mm]{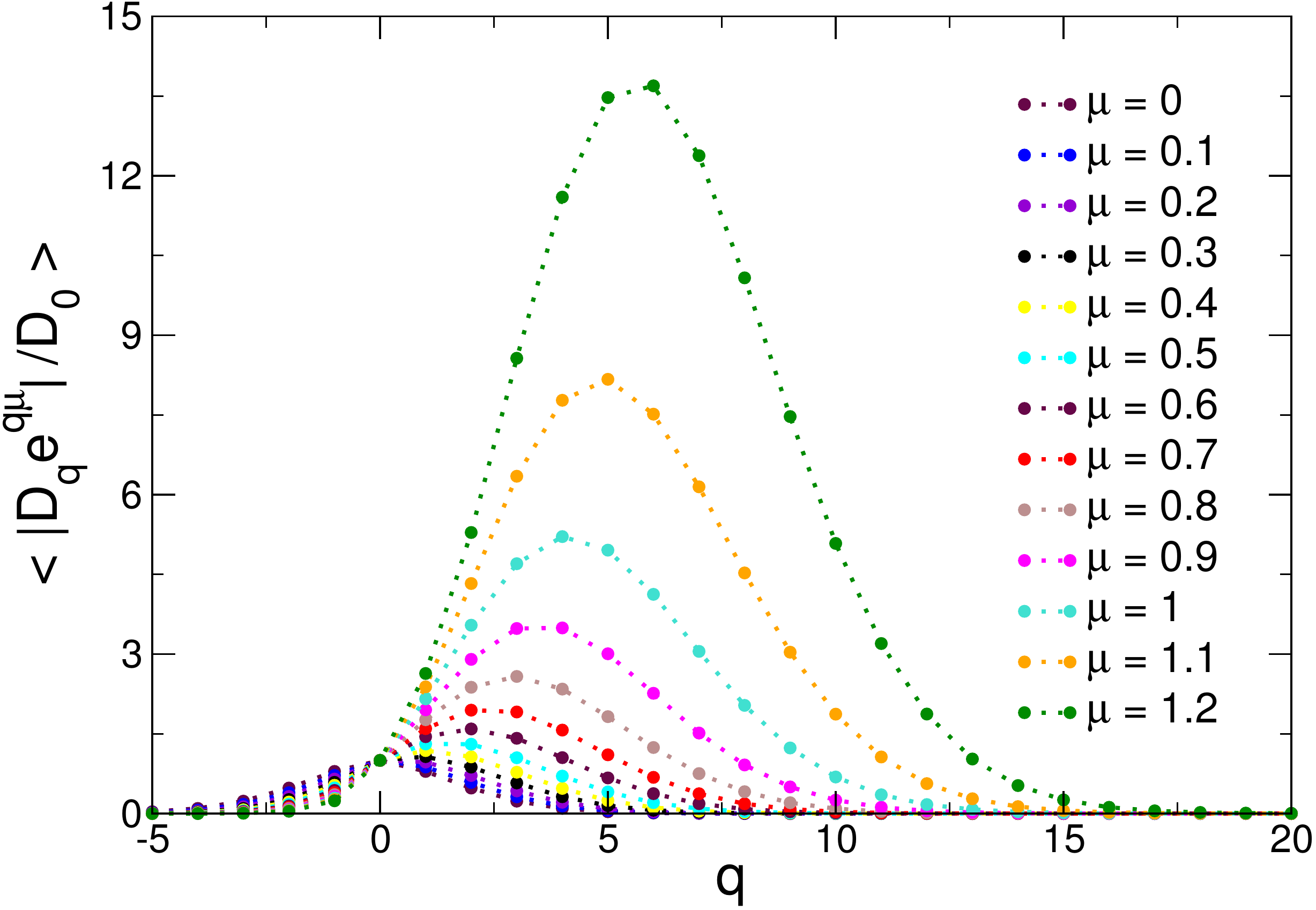}
\caption{Distribution of the coefficients in the fugacity series. In the lhs.~plot we 
show  $\langle | D_q | / D_0 \rangle$ versus $q$ at $\kappa = 0.001, \mu = 0$ for  
$16^3$ lattices at different values of temperature parameter $\tau$ (width of distribution increases with $\tau$). 
On the rhs.\ we show $\langle | e^{\mu q} D_q | / D_0 \rangle$ for $\kappa = 0.001, \tau = 0.183, 16^3$ 
for different values of $\mu$. Increasing $\mu$ increases the height of the maximum and shifts it to larger values of 
$q$.}
\label{gauss_CAP}
\end{figure}

In the lhs.~plot of Fig.~\ref{gauss_CAP} we show the expectation value $\langle | D_q | / D_0 \rangle$ 
versus $q$ at $\kappa = 0.001$ ($\mu = 0$)  for different values of $\tau$ on $16^3$ lattices. The distribution 
has a Gaussian-like shape, with the width of the distribution increasing with the temperature parameter $\tau$. 
This behavior is to be expected, since the width of the distribution is related to the particle number susceptibility 
which increases with $\tau$. The analysis shows that on the $16^3$ lattices for all values of $\tau$ we consider, 
the main contributions to the $\mu = 0$ fugacity series are taken into account for $q_l = -10$, $q_u = +10$. 
 
The chemical potential enters the fugacity series via the factor $e^{\mu q}$, shifting the $D_q$ that contribute 
to the fugacity expansion towards larger values of $q$. This is evident from the rhs.\ plot in Fig.~\ref{gauss_CAP},  
where we show $\langle | e^{\mu q} D_q | / D_0 \rangle$ versus $q$ for $\kappa = 0.001, \tau = 0.183$ for different 
values of $\mu$. For the range of chemical potential values considered here a reasonable choice for the truncated 
series would be $q_l = -5$, $q_u = +20$. The optimal truncation values $q_l$ and $q_u$ for all parameters
we consider were determined by systematically studying the relative error between the exact expression and the 
truncated series as a function of $q_l$ and $q_u$. Observables are computed as derivatives of $\ln Z(\mu)$ with 
respect to the parameters.

\subsection{Regular Taylor expansion (RTE)}

The observables we consider here can be obtained as derivatives of the logarithm of the partition function $Z(\mu)$. 
The simplest series approach is to expand this logarithm in a Taylor series 
\begin{equation}
\ln Z(\mu) \;  = \; \sum_{n = 0}^\infty  \frac{\mu^n}{n!}  \left(\! \frac{\partial}{\partial \mu}\! \right)^n \, \ln Z (\mu) \, \bigg|_{\mu = 0} \; .
\end{equation}
The coefficients of the series are the corresponding derivatives of $\ln Z(\mu)$ at $\mu = 0$, i.e., combinations of 
moments of the magnetization $M$. Since $Z(\mu)$ is an even function in $\mu$, the odd coefficients vanish in 
the expansion of $\ln Z(\mu)$. Again we obtain observables as derivatives of the series for $\ln Z (\mu)$.

\begin{figure}[t]
\includegraphics[height = 45mm]{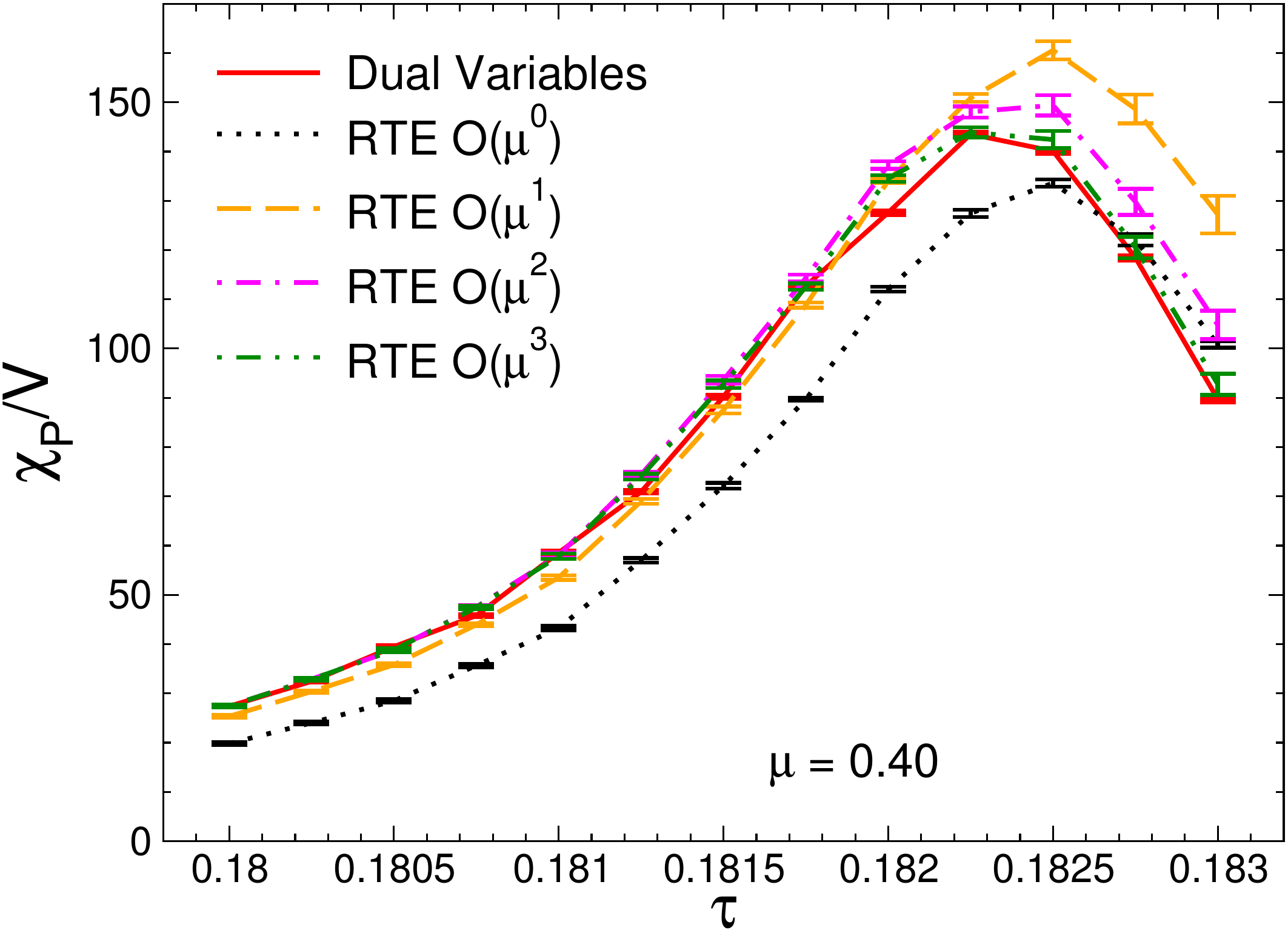}
\hspace{2mm} 
\includegraphics[height = 45mm]{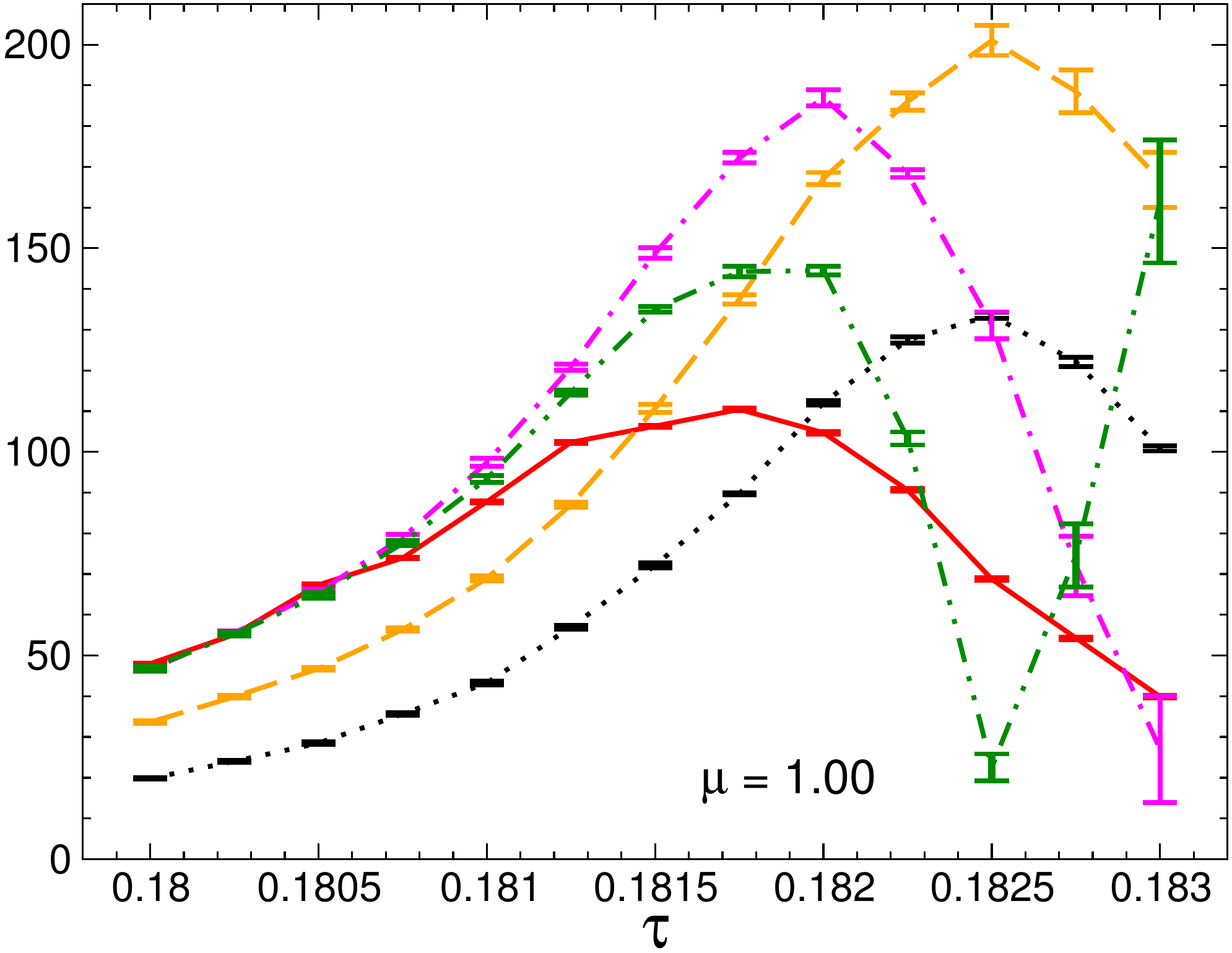}
\caption{Polyakov loop susceptibility obtained from the regular Taylor expansion RTE
for $\kappa = 0.001$, $\mu = 0.4$ (lhs.\ plot) and $\mu = 1.0$ (rhs.)
as a function of the temperature. Series results summed up to different orders in $\mu$ are shown 
and are compared to flux representation results.}
\label{fig:chiP_k0001_RTE_dual}
\end{figure}

It is important to understand that the coefficients $(\partial / \partial \mu )^n \, \ln Z (\mu) \, \mid_{\mu = 0}$ correspond 
to (higher) susceptibilities which fluctuate or even may diverge at transition points. They are multiplied with $\mu^n$ and thus
for larger values of $\mu$ are amplified and introduce fluctuations in the Taylor series. This property can be observed in 
Fig.~\ref{fig:chiP_k0001_RTE_dual} where we show the Taylor expansion results for the Polyakov loop susceptibility
accumulated up to different orders in $\mu$  ($16^3$ lattices with a statistics of $10^6$ 
configurations, $\kappa = 0.001$, $\mu = 0.4$ (lhs.\ plot) and $\mu = 1.0$ (rhs.)). 

The lhs.~plot demonstrates that for $\mu = 0.4$ the approximation of the reference results from the dual simulation improves 
when more terms are added in the Taylor expansion. For $\mu = 1.0$ (rhs.\ plot) this is no longer the case and it is obvious
that including the higher terms does not improve the result, but instead adds fluctuations. 

\subsection{Improved Taylor expansion (ITE)}

We now consider a second type of Taylor series, which we refer to as the ''improved Taylor expansion'', 
where the logarithm of the partition sum is expanded 
in a double Taylor series in the parameters 
$ \rho = e^{\mu} - 1 $ and $\bar{\rho} =  e^{- \mu} - 1 $ 
(which in the limit $\mu \rightarrow 0$ reduces to an expansion in $\mu$). 
A part of the motivation for this type of expansion is to capture some of the features of the fugacity 
expansion, which in the case of QCD would lead to a finite Laurent series, whereas the regular 
Taylor expansion gives rise to an infinite series.

For the ITE the Boltzmann factor is organized as follows, 
\begin{equation}
e^{- S_{\mu}} \; = \; e^{- S_0} \; e^{\, \rho R + \bar{\rho} \bar{R}},
\label{ITE1}
\end{equation}
where $S_\mu$ is the action at finite $\mu$ and $S_0$ the action at $\mu = 0$. $R$ and $\bar{R}$ are the terms 
multiplied with the factors $e^{\pm \mu}$. In full QCD these are the temporal hopping terms for the fermions, in the 
$\mathds{Z}_3$ spin model they are given by $R = \kappa M$ and $\bar{R} = \kappa M^*$. Thus the second term
in (\ref{ITE1}) simply removes the terms $R$ and $\bar{R}$ from the action and reinstates them after multiplication with
$e^{\, \pm \mu}$. 
The partition sum now has the obvious expansion 
\begin{eqnarray}
Z(\mu) & \, = \,& \sum_{\{ P \}} e^{- S_{\mu}} \, =  \sum_{\{ P \}} e^{- S_0} \; e^{\, \rho R } \; e^{\, \bar{\rho} \bar{R}}
\label{ITE2} \\
& \, = \, &  \sum_{\{ P \}} e^{- S_0} \!\!\!\sum_{n,m=0}^\infty \! \frac{\rho^n}{n!} \frac{\bar{\rho}^{\,m}}{m!} \, R^n \, \bar{R}^m \, = \, 
Z(0) \!\! \sum_{n,m=0}^\infty \! \frac{\rho^n}{n!} \frac{\bar{\rho}^{\,m}}{m!} \, \left\langle R^n \, \bar{R}^m \right\rangle_0  ,
\nonumber
\end{eqnarray}
where $\langle ... \rangle_0$ is the expectation value of the $\mu = 0$ theory. In (\ref{ITE2}) we use the language of the spin 
system for the presentation of the ITE. For QCD the sum $\sum_{\{P\}}$ over spin configurations has to be replaced
by the path integral $\int {\cal D}[U,\psi, \overline{\psi}]$ over the gauge and fermion degrees of freedom. The logarithm of the 
partition function for the evaluation of observables is obtained by a further double-expansion in $\rho$ and $\bar{\rho}$ and 
observables evaluated by subsequent derivatives. 

The expansion coefficients $\left\langle R^n \, \bar{R}^m \right\rangle_0$ of the ITE 
have a structure different from the terms in the regular Taylor series but their evaluation in 
full QCD has the same numerical cost as the coefficients of the regular Taylor expansion RTE.
When $\rho = e^\mu - 1$ and $\bar{\rho} = e^{-\mu} - 1$  are expanded in $\mu$ one of 
course gets back the RTE and the coefficients of the ITE are sums of coefficients of the RTE. 
Thus one can view the improved Taylor expansion also as a partially resummed conventional
Taylor expansion.

\begin{figure}[t]
\includegraphics[height = 47mm]{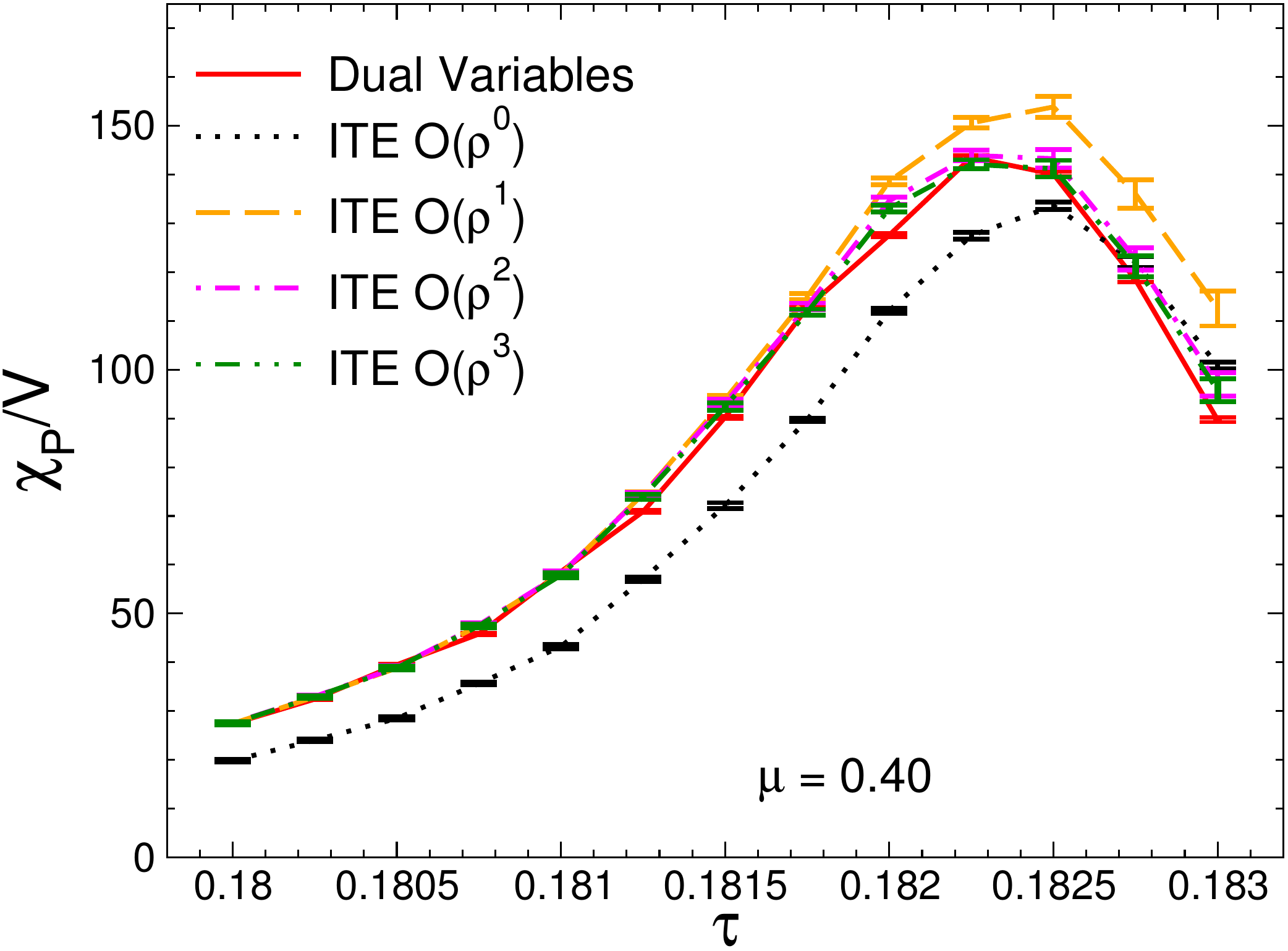}
\hspace{1mm}
\includegraphics[height = 47mm]{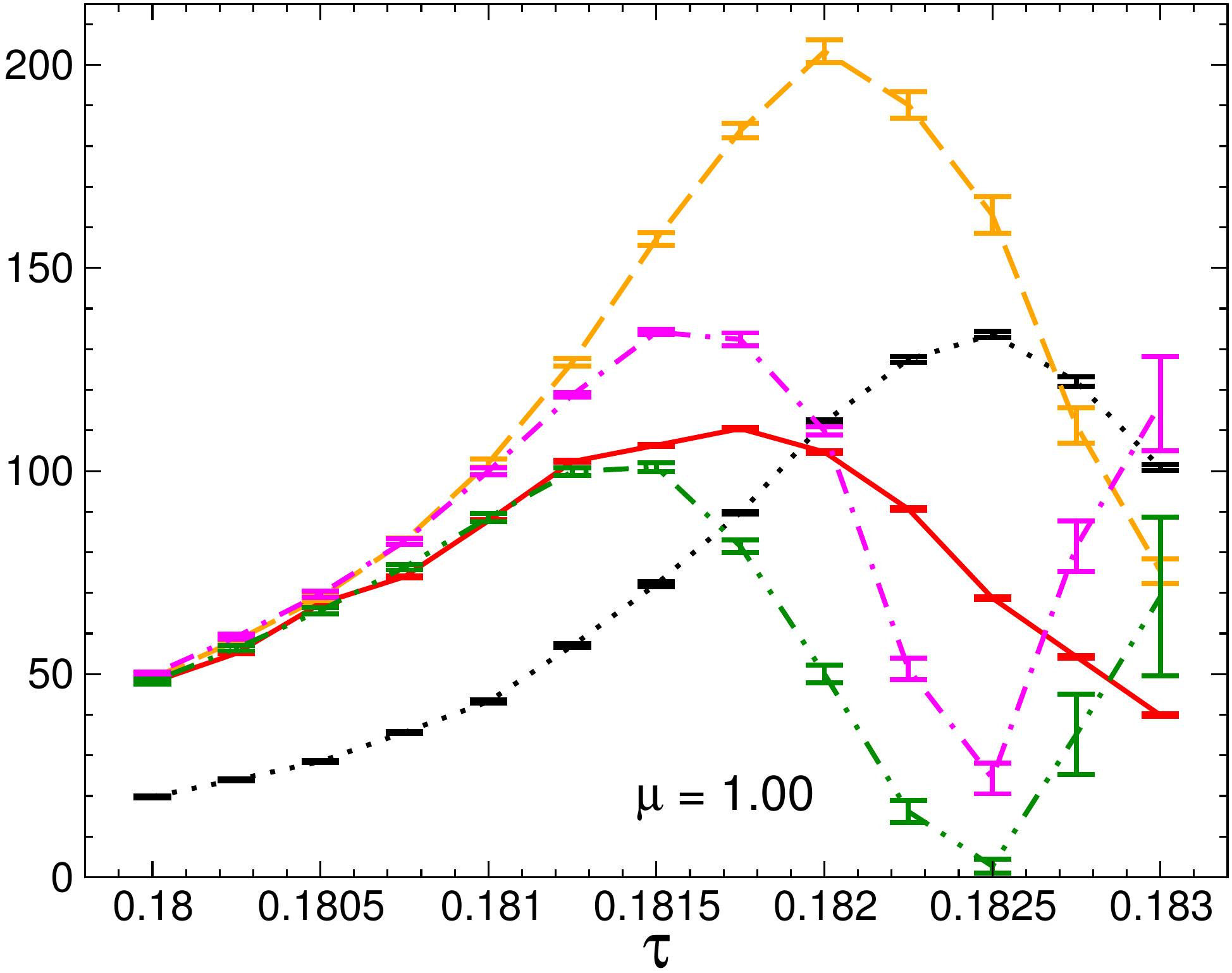}
\caption{Polyakov loop susceptibility obtained from the improved Taylor expansion ITE for 
$\kappa = 0.001$, $\mu = 0.4$ (lhs.\ plot) and $\mu = 1.0$ (rhs.)
as a function of the temperature. Results up to third order of $\rho$ are shown 
and are compared to flux representation results.}
\label{fig:chiP_k0001_ITE_dual}
\end{figure}

Similar to the case of the regular Taylor expansion, in Fig.~\ref{fig:chiP_k0001_ITE_dual} we analyze the
buildup of the series expansion result when adding higher orders in the expansion. 
Fig.~\ref{fig:chiP_k0001_ITE_dual} shows the Polyakov loop susceptibility at $\kappa = 0.001$ as a 
function of the temperature for $\mu = 0.4$ (lhs.\ plot) and for $\mu = 1.0$ (rhs.). We display the results
from different truncation orders of the ITE (the order of a term is defined as the sum of the orders of 
$\rho$- and $\bar{\rho}$-factors), and compare them to the reference results from the dual simulation.

As for the RTE, also the ITE quickly converges towards the dual results for $\mu = 0.4$. For $\mu = 1.0$ we no longer see
overall convergence for all $\tau$ values and again observe that higher terms fluctuate in $\tau$. However, a 
comparison with Fig.~\ref{fig:chiP_k0001_RTE_dual} shows that at the same numerical cost the third order ITE converges
to the dual results in a larger $\tau$-interval than the corresponding third order RTE.
 
\section{Direct comparison of all three expansion techniques}

Having introduced the three series expansions and having discussed some of their features, we now come to a systematic
comparison of the three expansions and the analysis of their convergence to the reference data from the dual simulation. 

We compare the expansions for two values of the inverse mass parameter: For $\kappa = 0.01$ (''light quarks'')
and for $\kappa = 0.001$ (''heavy quarks''). As we will see below (and as is expected for general reasons), the sign problem is
more severe for the light quarks (compare also Fig.~\ref{CAP}), and thus for the light quarks $\kappa = 0.01$ we expect the series 
to fail already at smaller $\mu$ than in the heavy quark case $\kappa = 0.001$. Consequently, at $\kappa = 0.01$ 
we show results for chemical potential values of $\mu = 0.2, 0.6, 0.8, 1.0$ and $1.2$, while for $\kappa = 0.001$ we can go
to larger values and use $\mu = 0.2, 0.8, 1.0, 2.2$ and $3.0$ for the comparison. 

For the fugacity expansion we use the procedure discussed in Section 3.1 for determining the optimal truncation
values $q_l$ and $q_u$. The regular Taylor expansion RTE and the improved Taylor expansion ITE are both compared here 
with terms up to third order included, such that the numerical cost is exactly the same for the two series (which is negligible here, 
but in a full QCD simulation is of course a key issue). The coefficients for all three series expansions were computed in  
$\mu = 0$ simulations with a statistics of $4\times10^7$ configurations, while for the reference data from dual simulations $10^6$ 
configurations were used. The error bars we show are statistical errors determined with the jackknife method.

\begin{figure}[p]
\hspace*{8mm} \includegraphics[height=38mm,width=50mm]{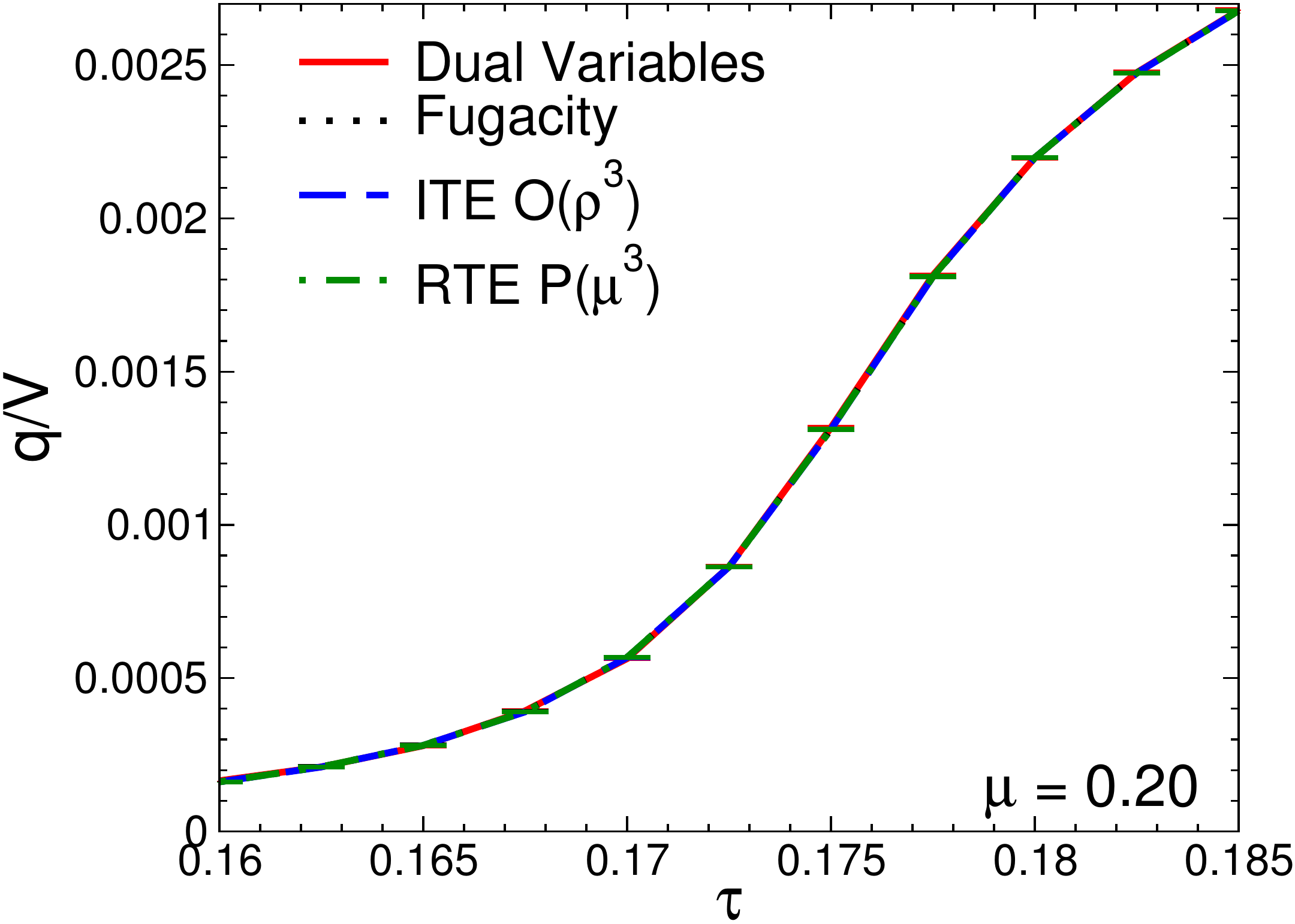}  
\hspace*{10mm}   \includegraphics[height=38mm,width=50mm]{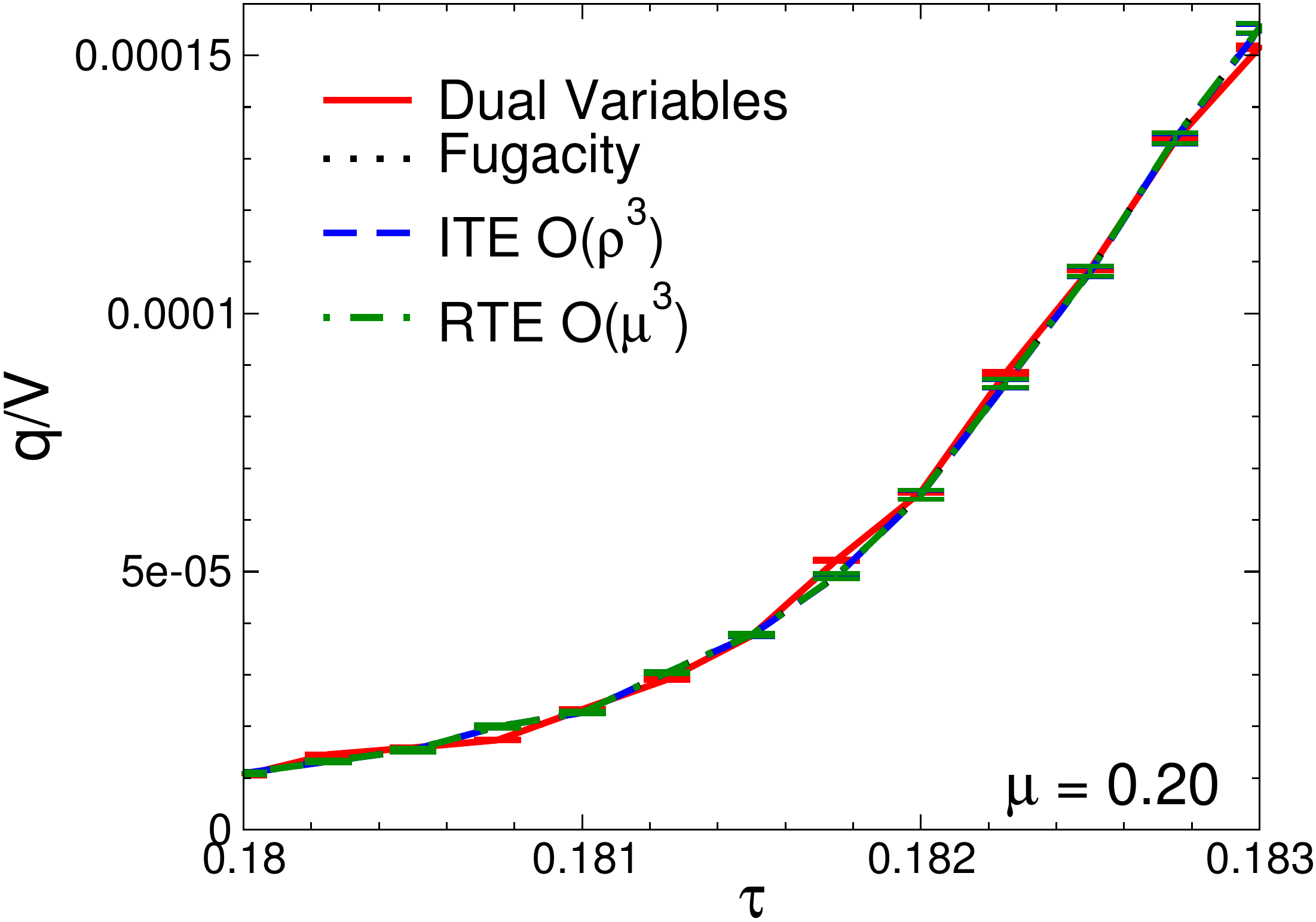}

\hspace*{9mm} \includegraphics[height=38mm,width=49mm]{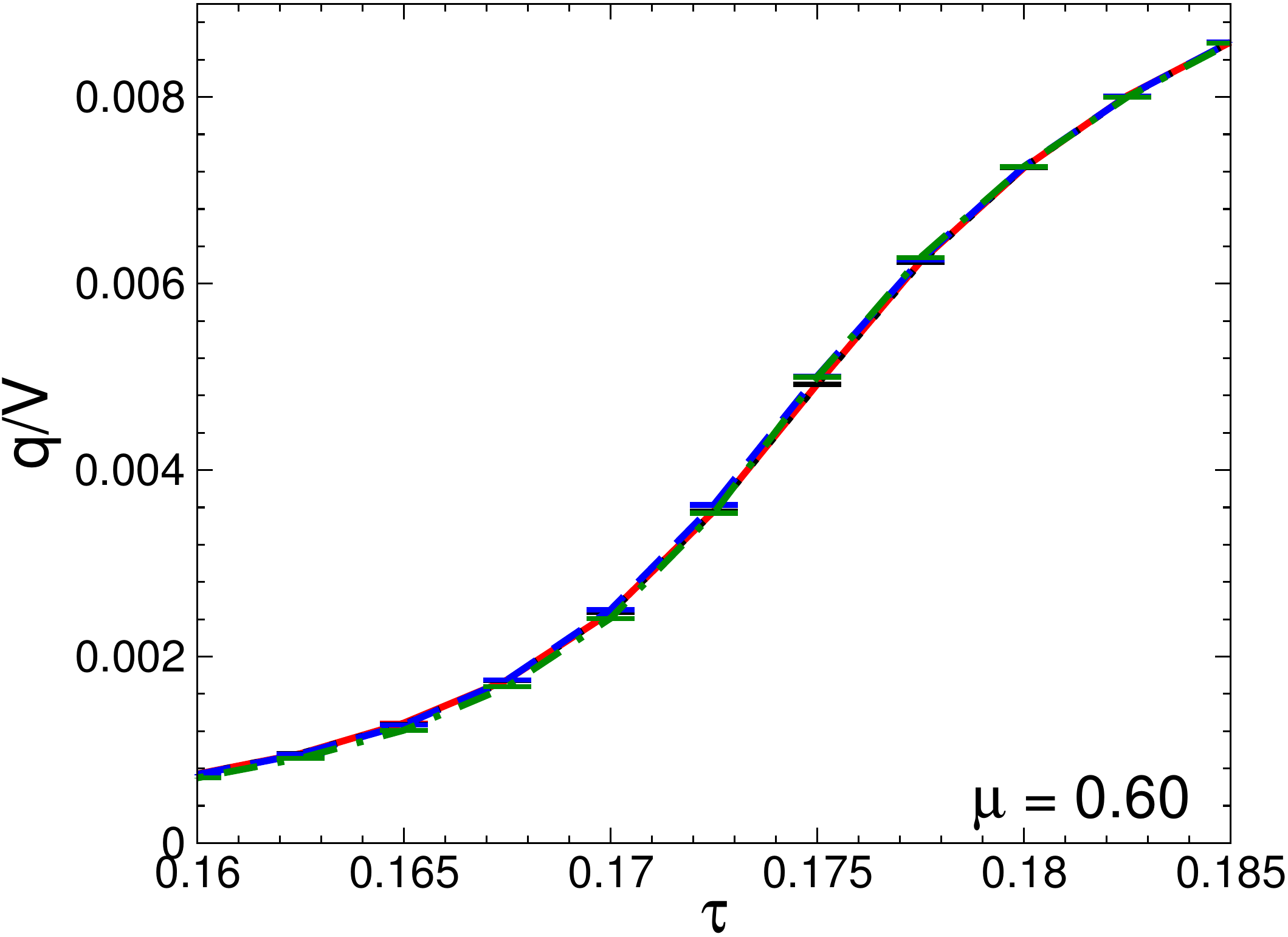}
\hspace*{11.1mm}   \includegraphics[height=38mm,width=49mm]{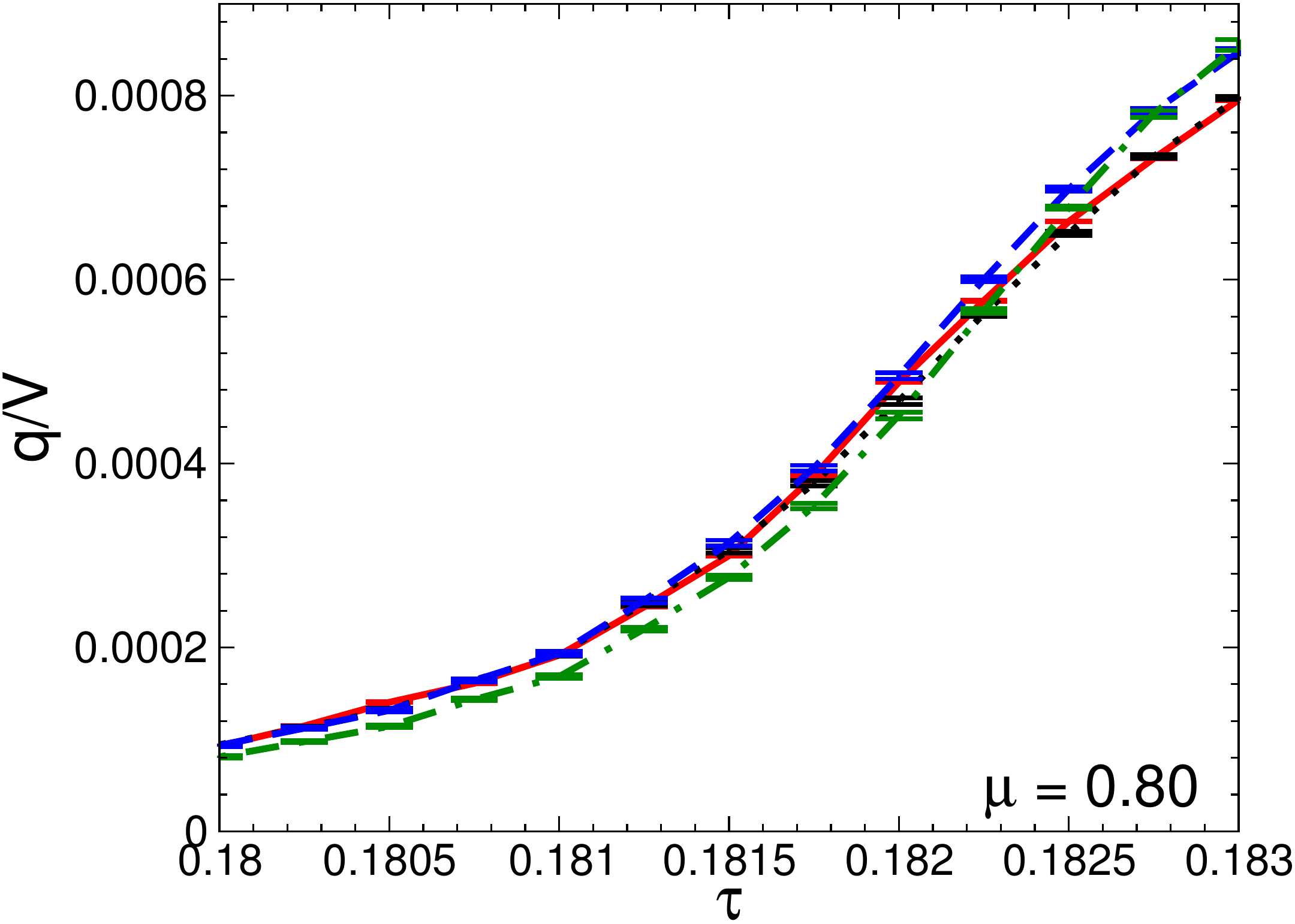}

\hspace*{9mm} \includegraphics[height=38mm,width=49mm]{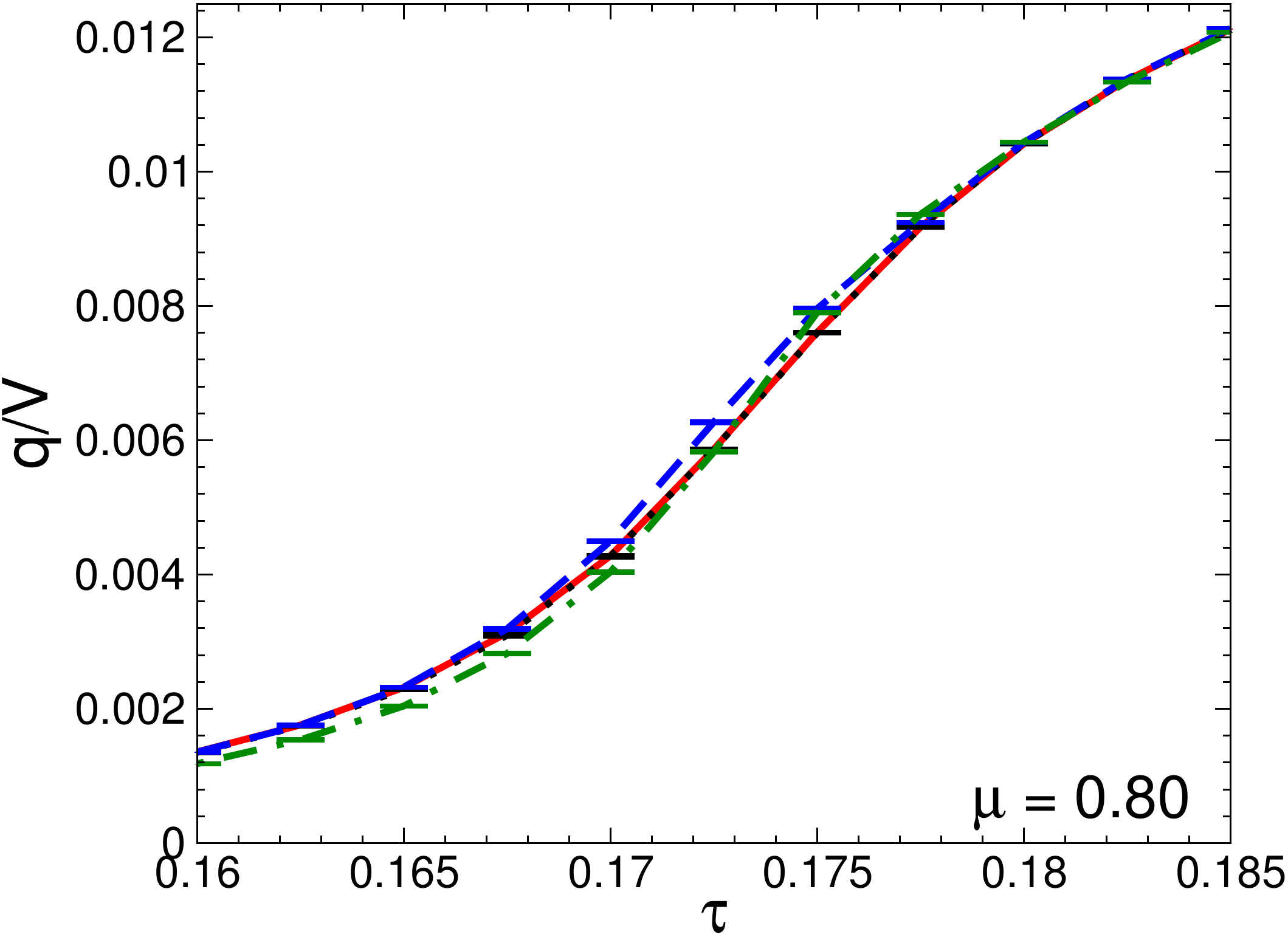}
\hspace*{11.1mm} \includegraphics[height=38mm,width=49mm]{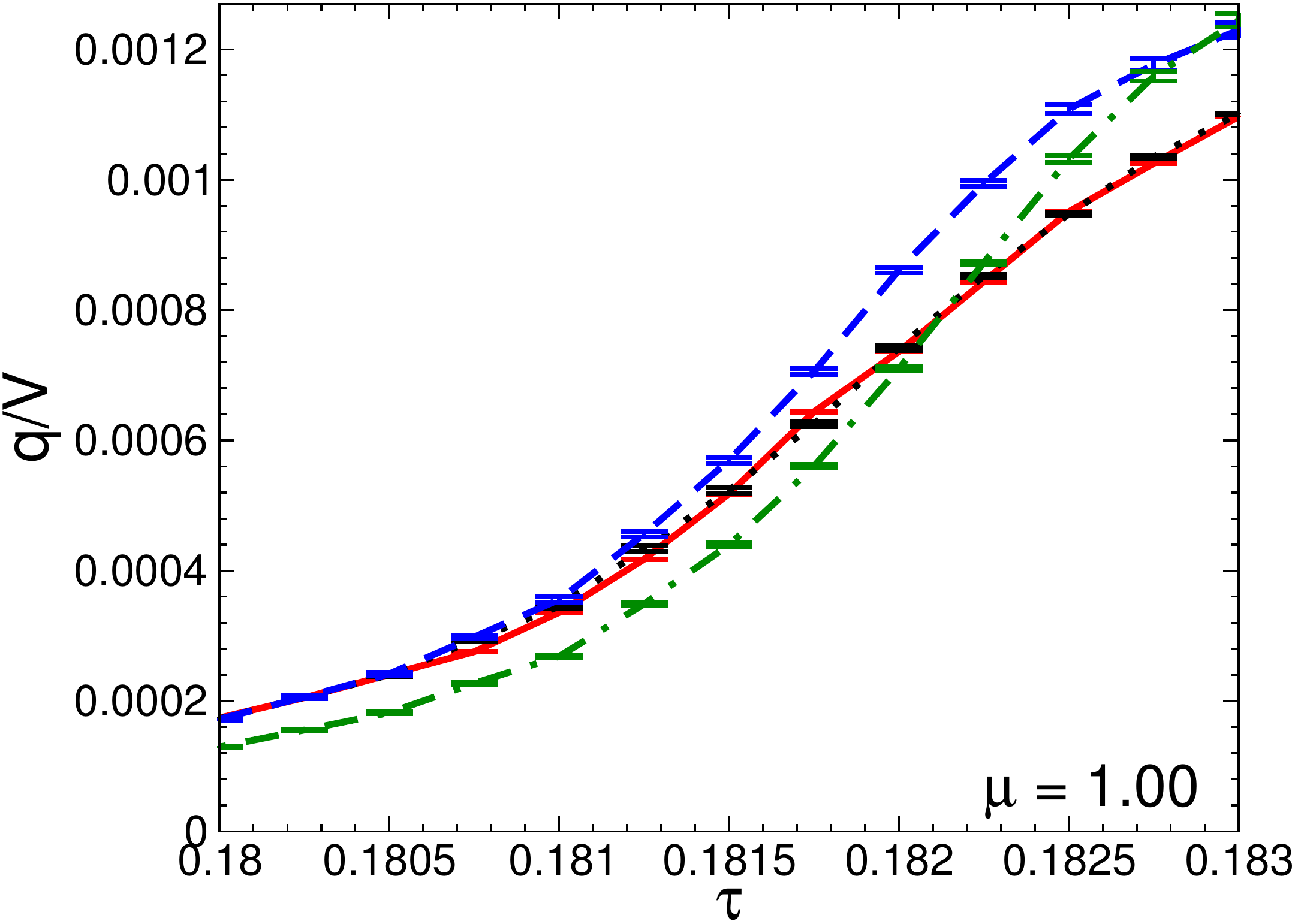}

\hspace*{9mm} \includegraphics[height=38mm,width=49mm]{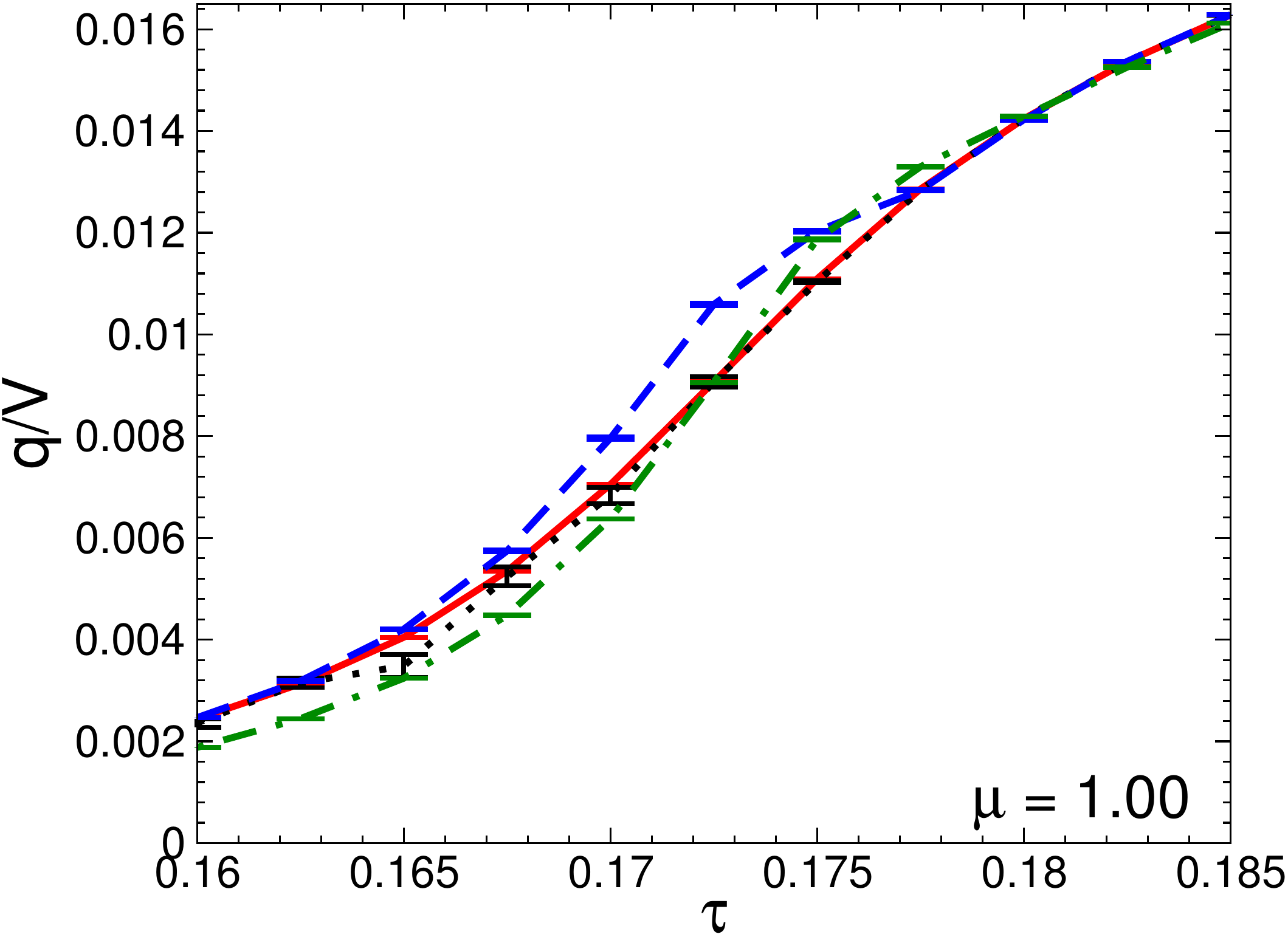}
\hspace*{12.5mm} \includegraphics[height=38mm,width=48mm]{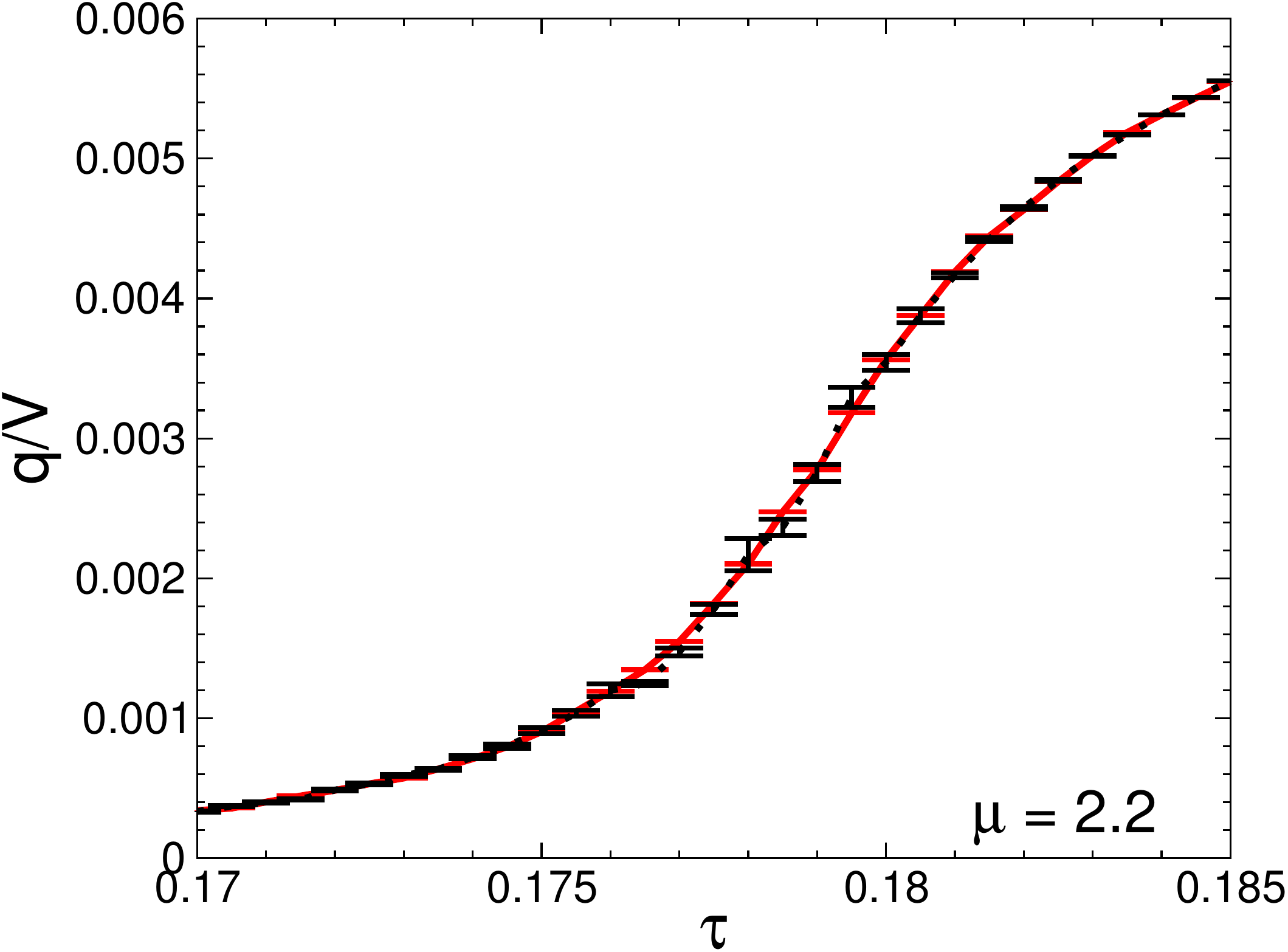}

\hspace*{9.1mm} \includegraphics[height=38mm,width=48.5mm]{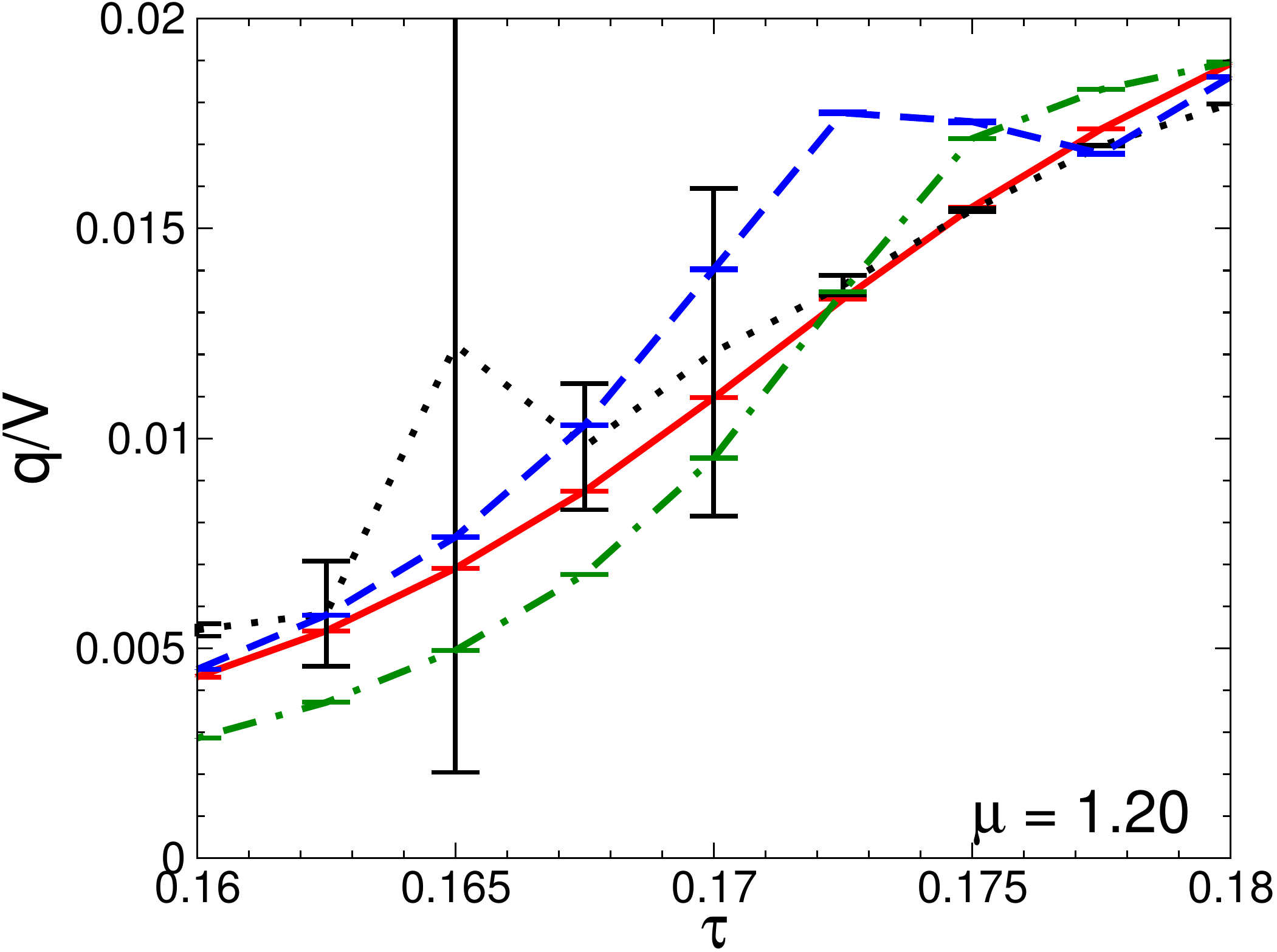}
\hspace*{12.9mm} \includegraphics[height=38mm,width=48mm]{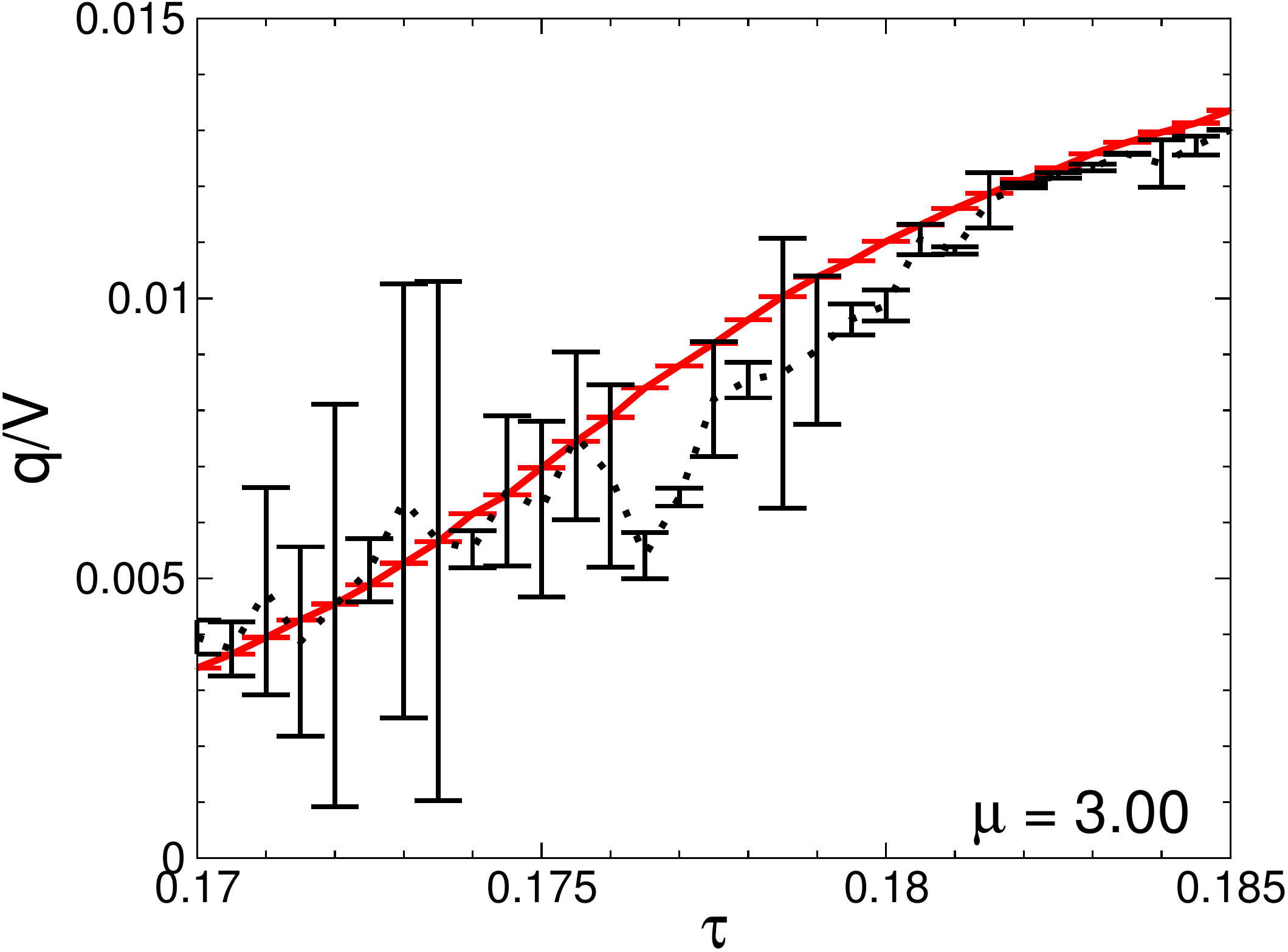}

\caption{Comparison of the particle number $q$ at $\kappa = 0.01$  (lhs.) and $ \kappa = 0.001$ (rhs.) 
from fugacity expansion, RTE, ITE and the dual simulation for different values of the chemical potential $\mu$. 
\label{q_vergl}}
\end{figure}

\begin{figure}[p]
\hspace*{8mm} \includegraphics[height=38mm,width=50mm]{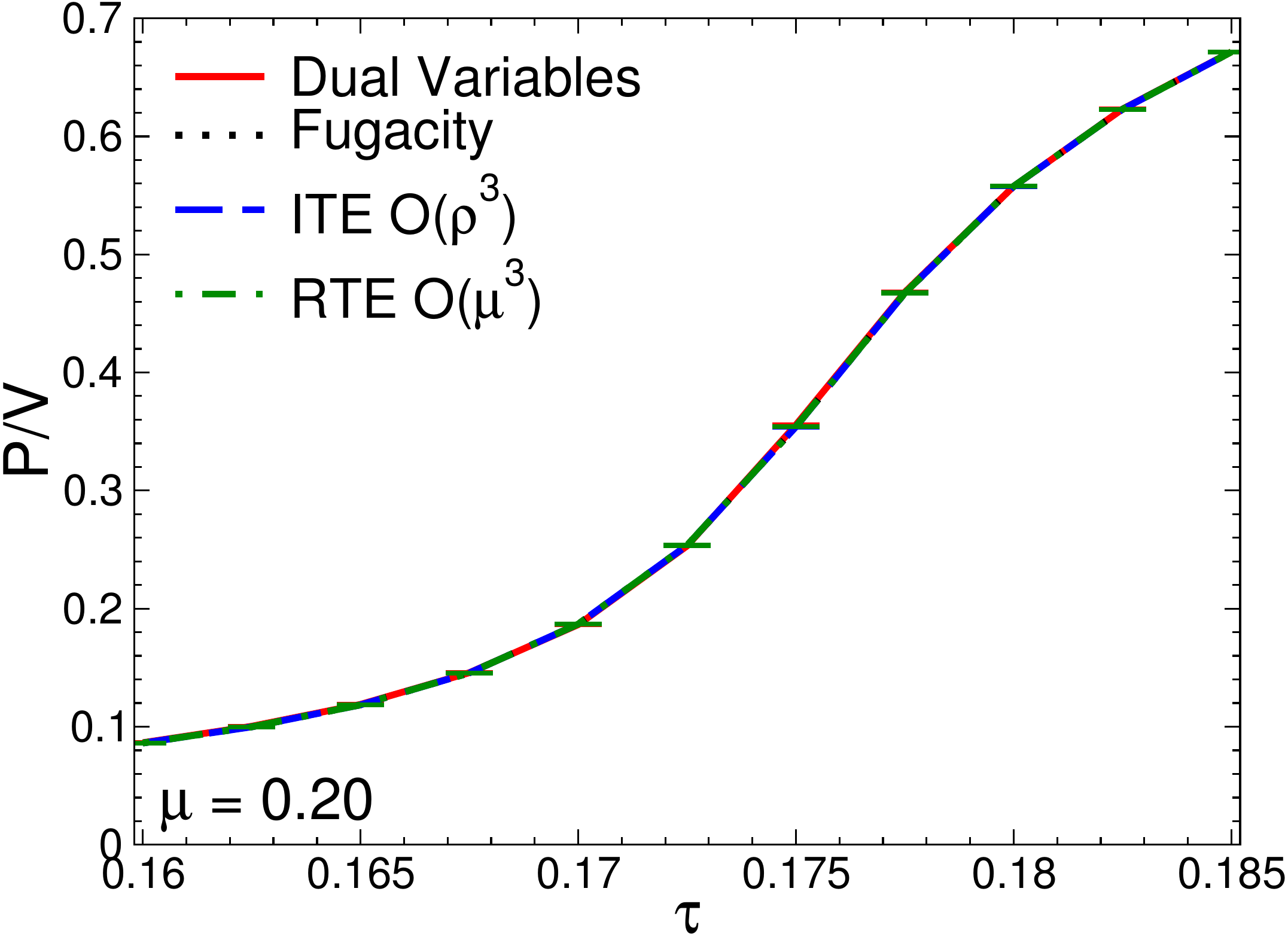}  
\hspace*{10mm}\includegraphics[height=38mm,width=50mm]{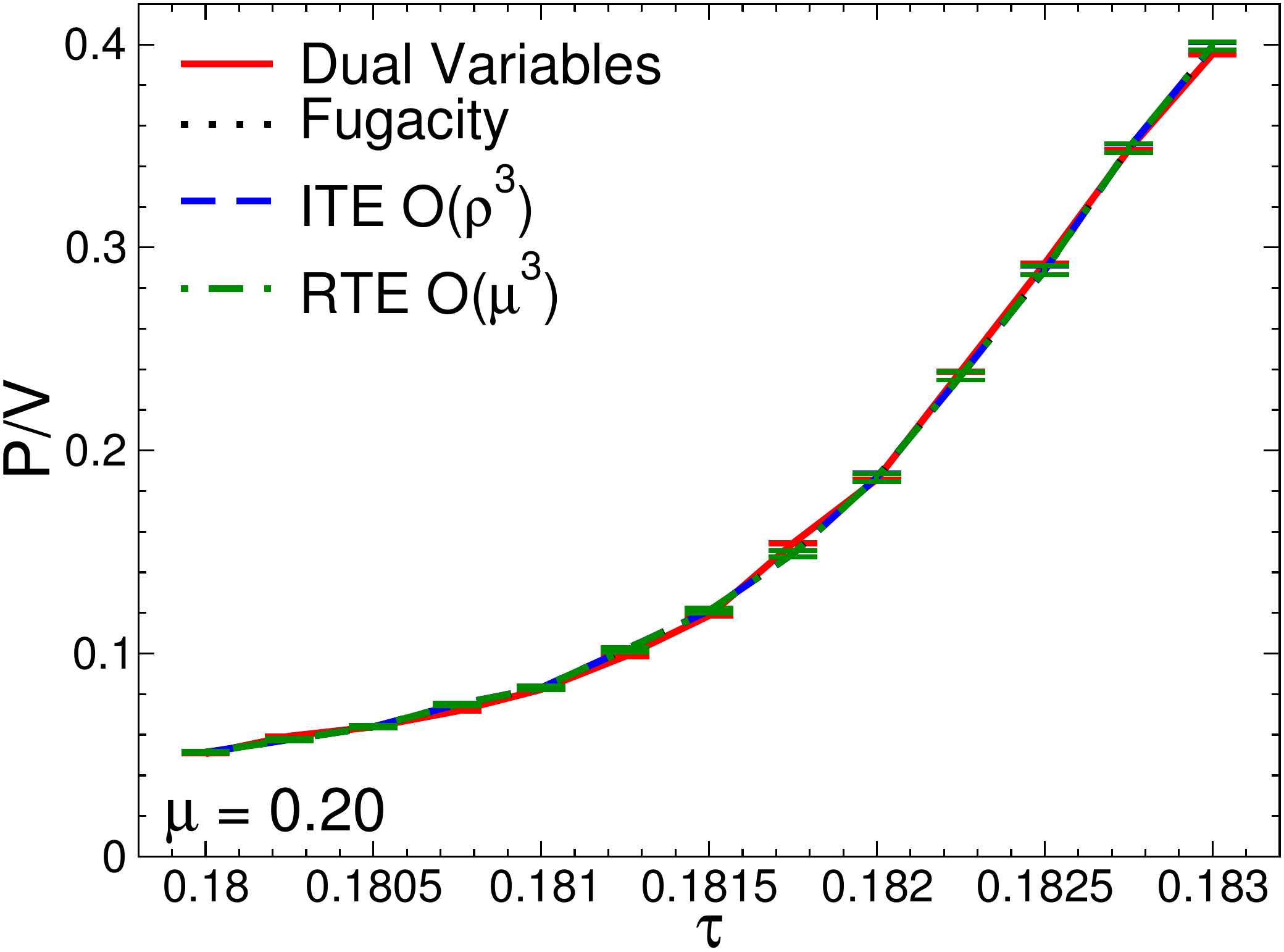}
 
\hspace*{9mm}\includegraphics[height=38mm,width=49.5mm]{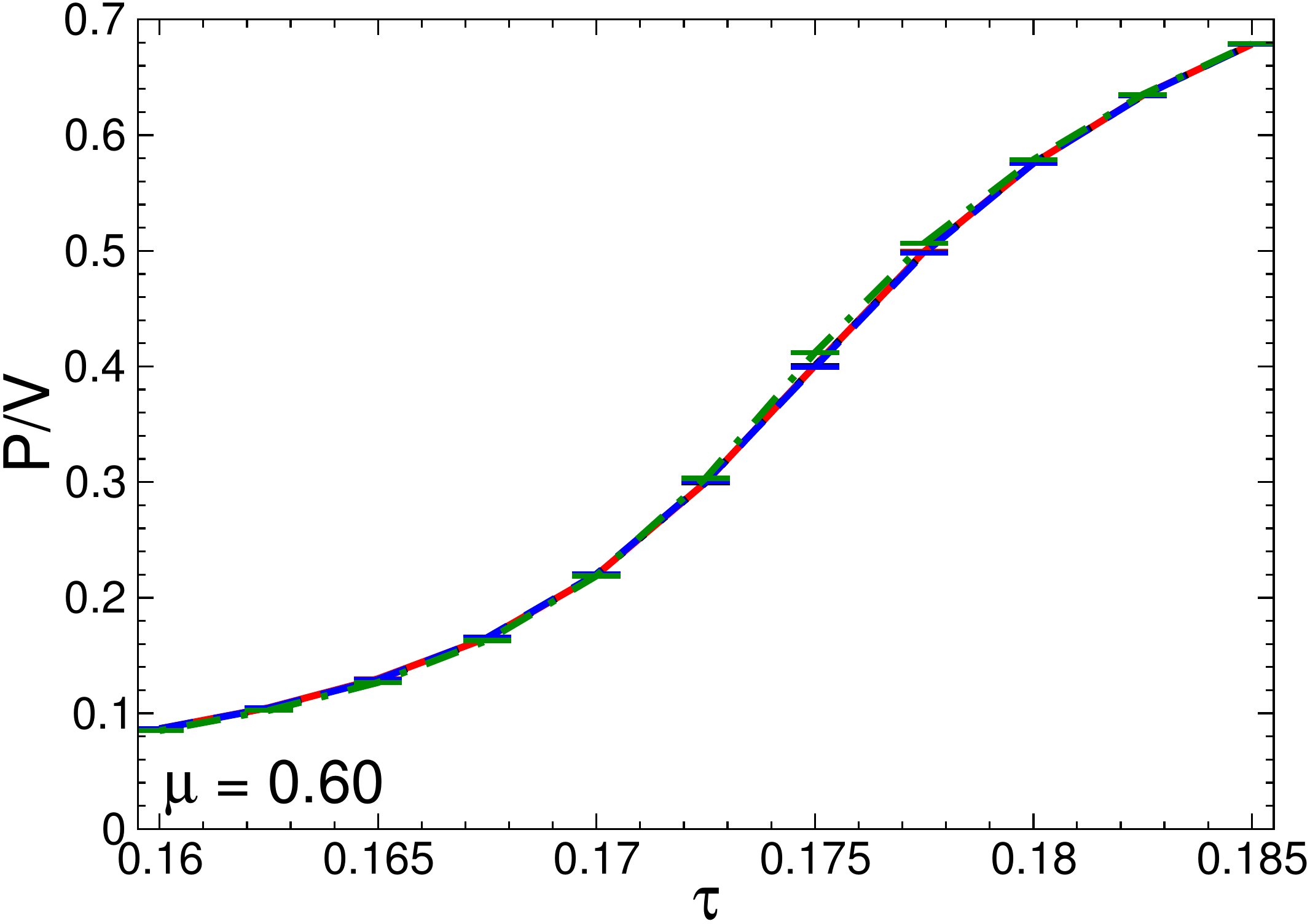}
\hspace*{10.8mm}\includegraphics[height=38mm,width=50mm]{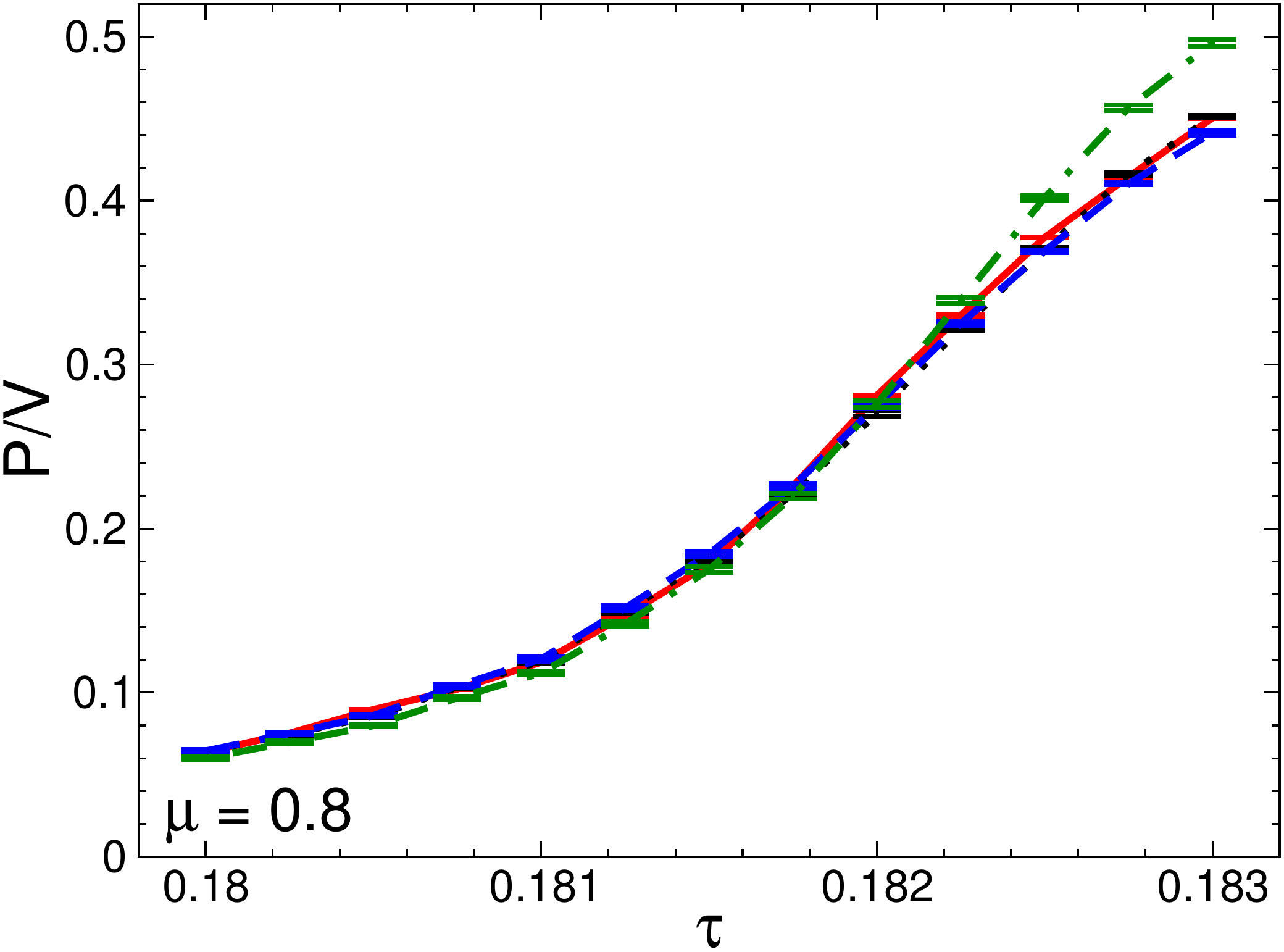}
 
\hspace*{9mm}\includegraphics[height=38mm,width=49.5mm]{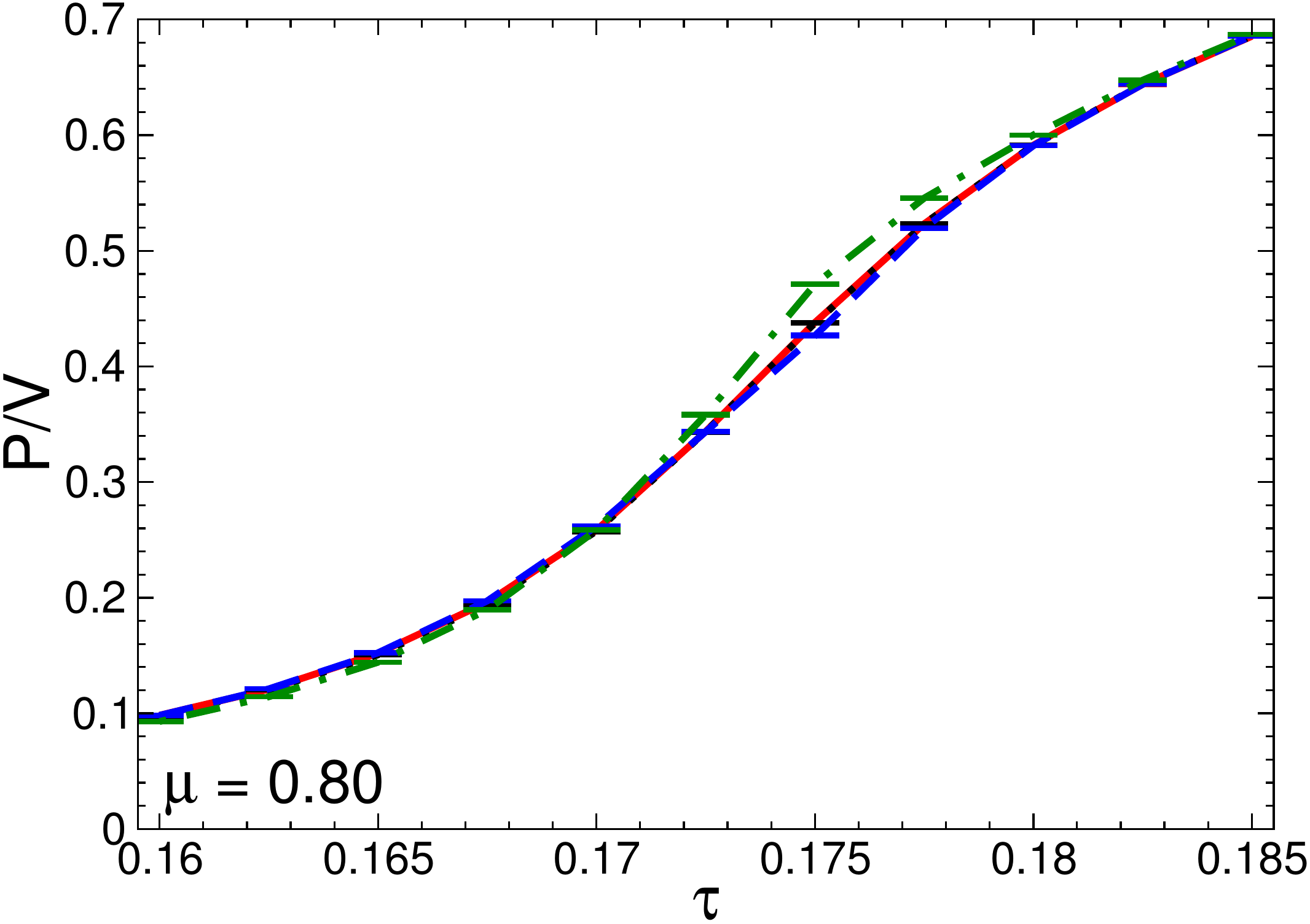}
\hspace*{10.8mm}\includegraphics[height=38mm,width=50mm]{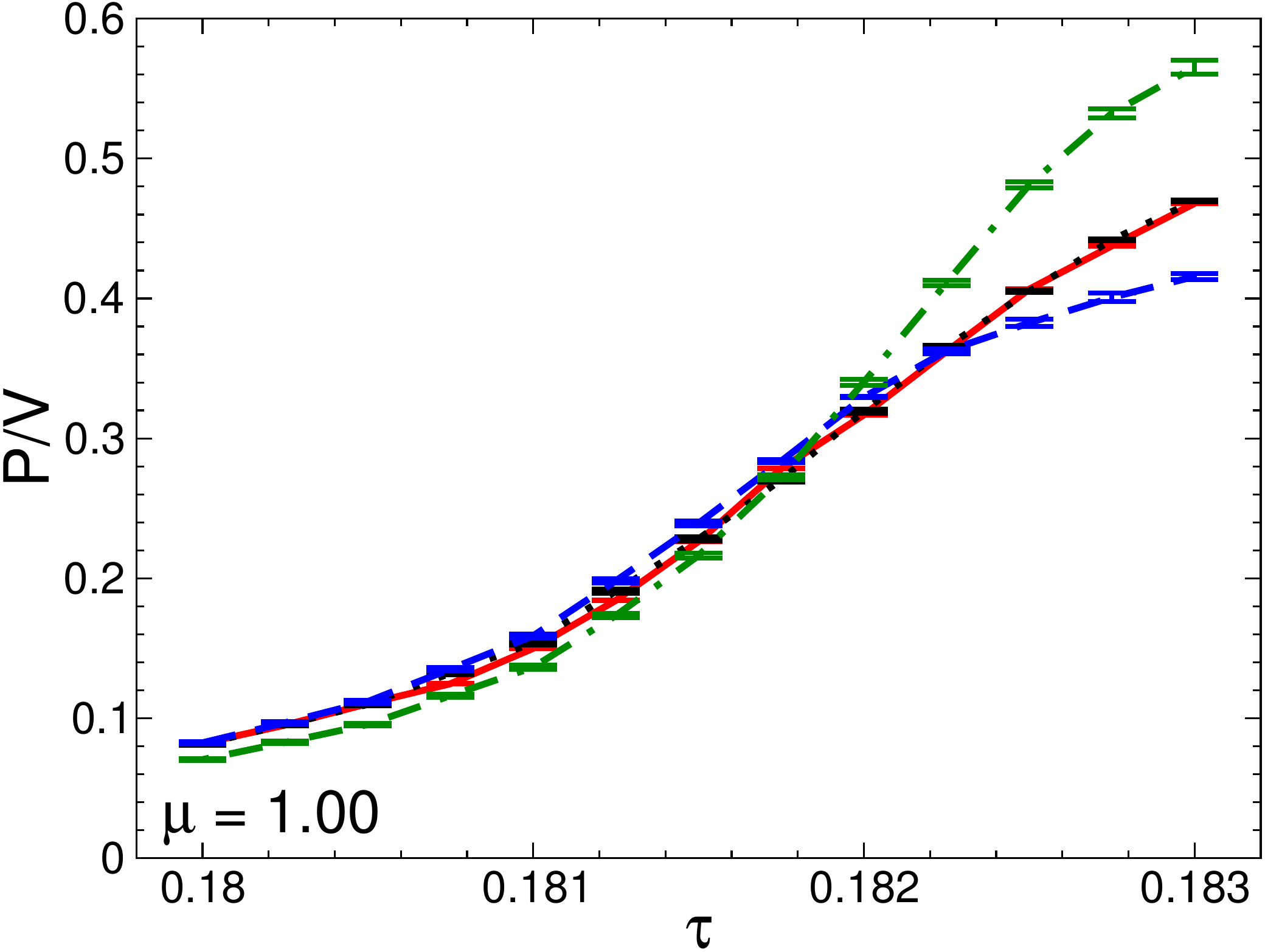}
 
\hspace*{9mm}\includegraphics[height=38mm,width=49.5mm]{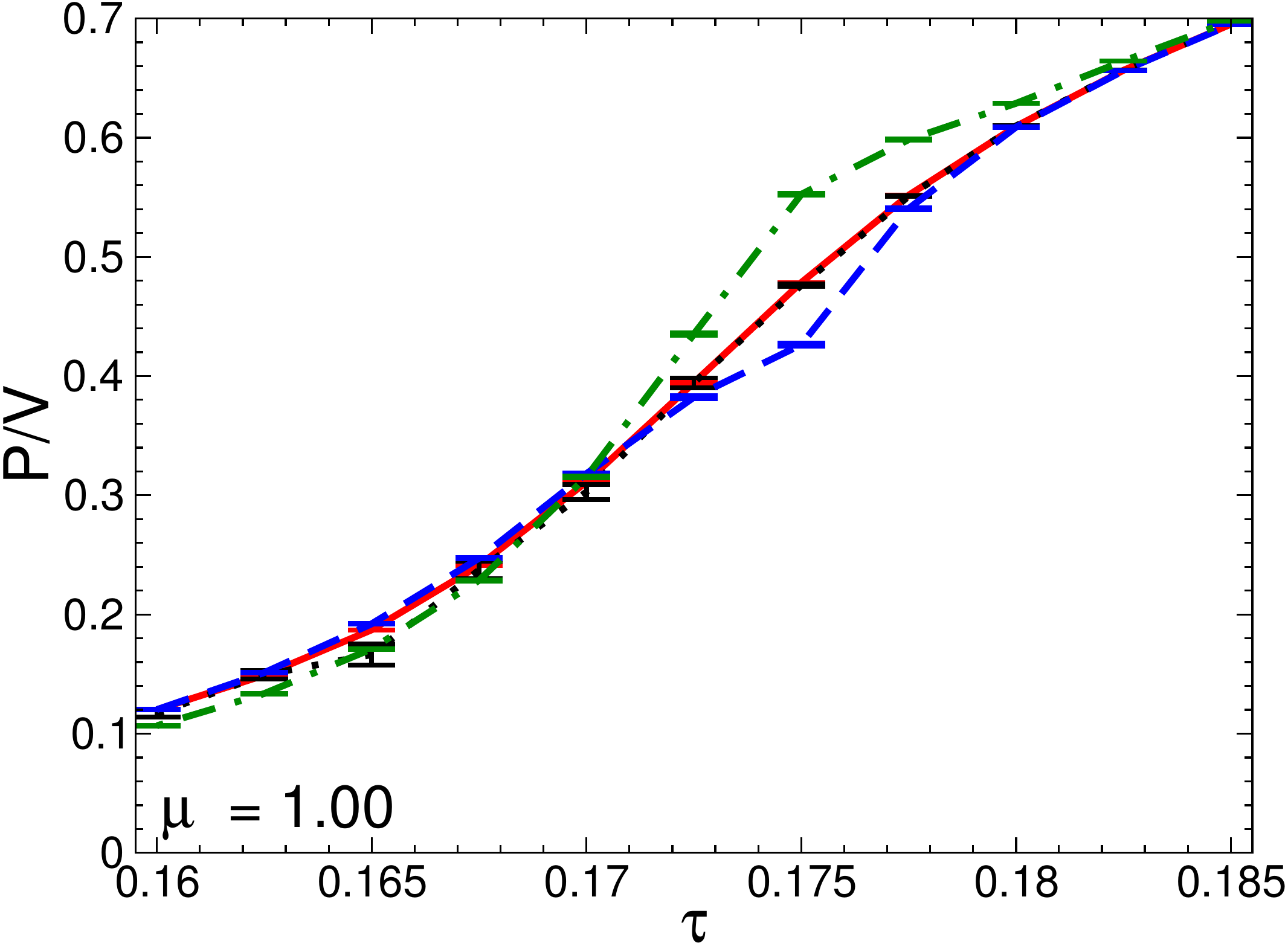}
\hspace*{10.8mm}\includegraphics[height=38mm,width=50mm]{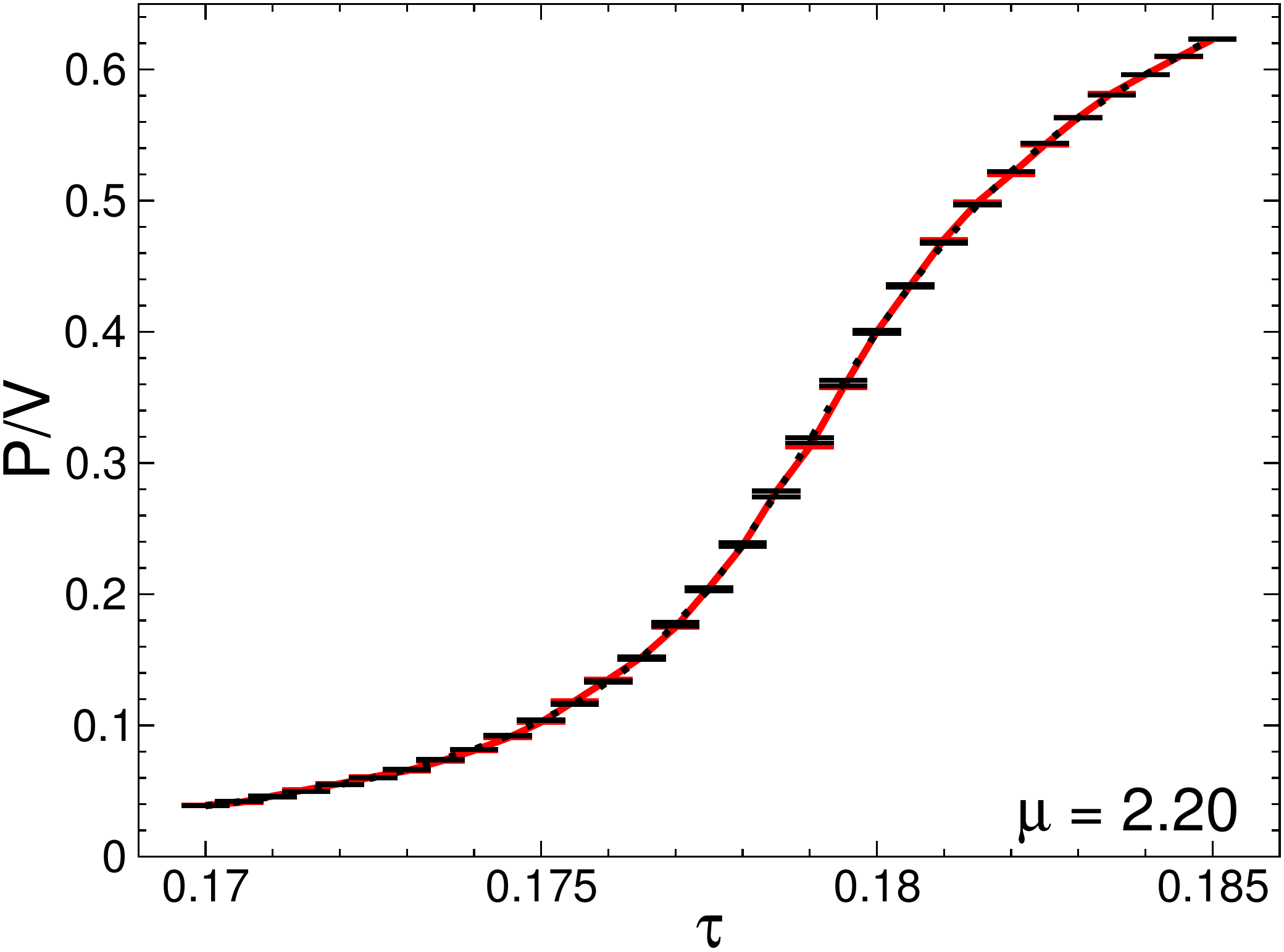}
 
\hspace*{9mm}\includegraphics[height=38mm,width=50mm]{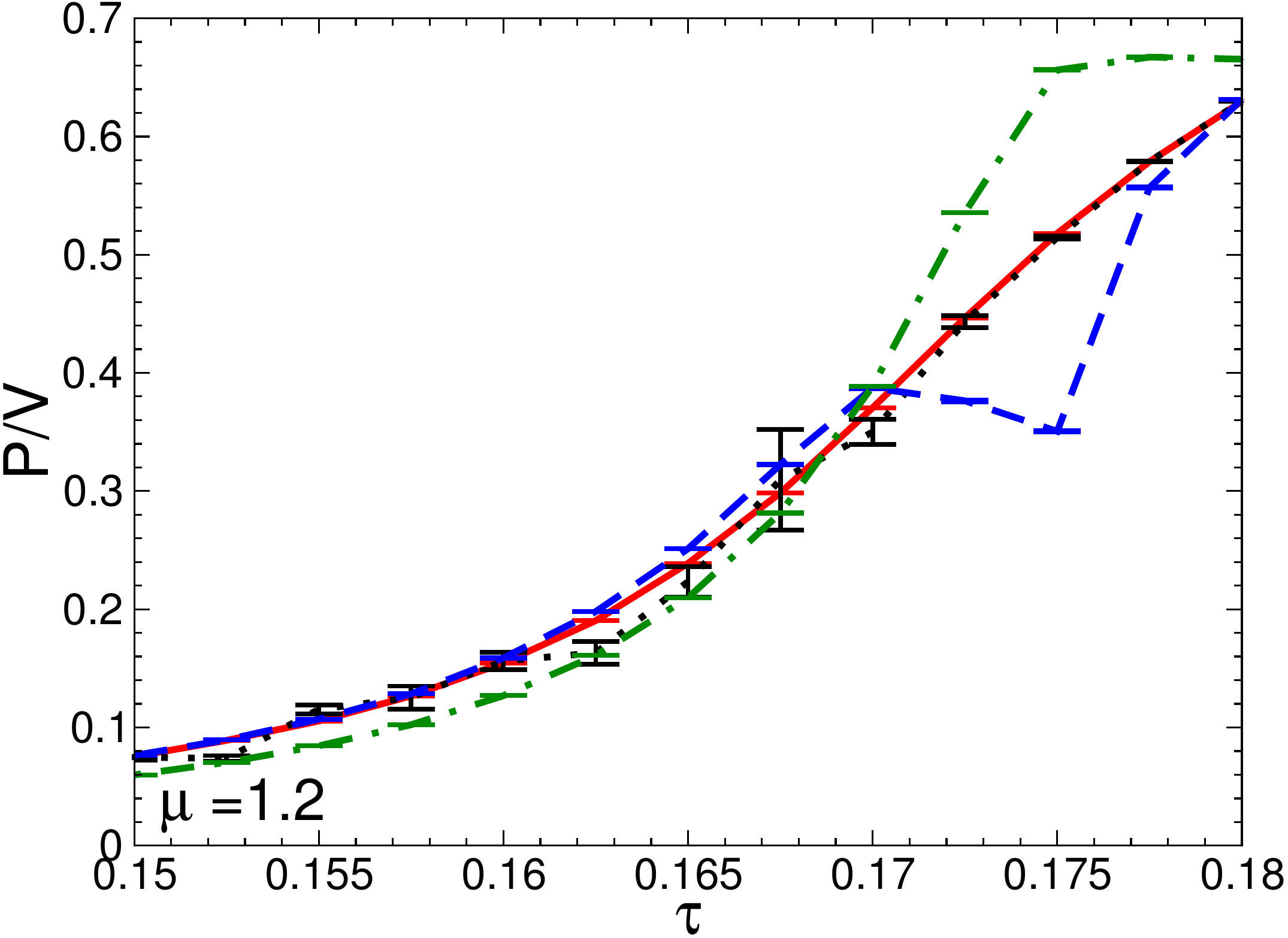}
\hspace*{10.4mm}\includegraphics[height=38mm,width=50mm]{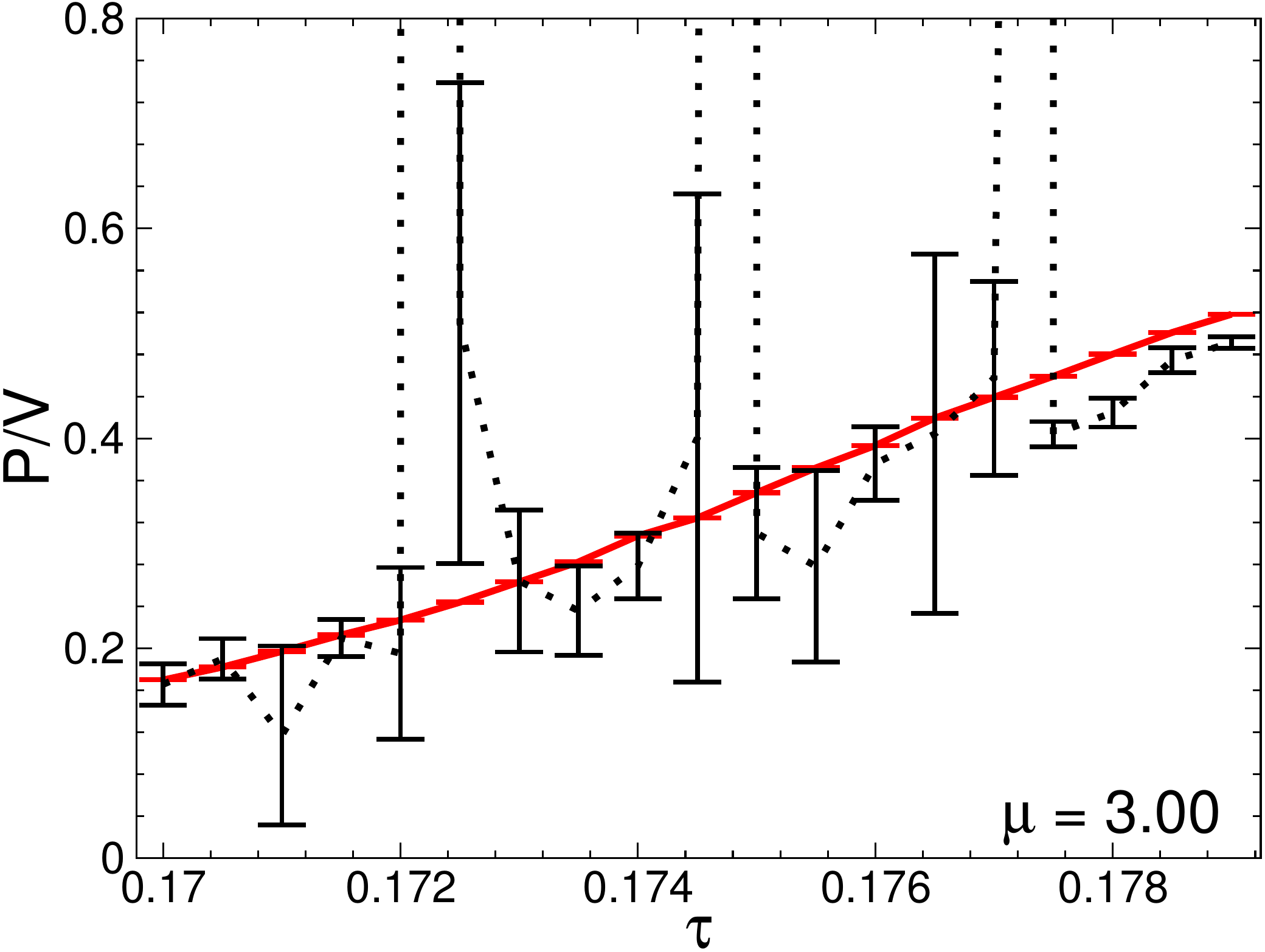}

\caption{The Polyakov loop $P$ at $ \kappa = 0.01$  (lhs.) and $ \kappa = 0.001$ (rhs.) 
from fugacity expansion, RTE, ITE and from the dual simulation for different 
values of the chemical potential $\mu$.
\label{P_vergl}}
\end{figure}

\begin{figure}[p]

\hspace*{8mm}\includegraphics[height=38mm,width=50mm]{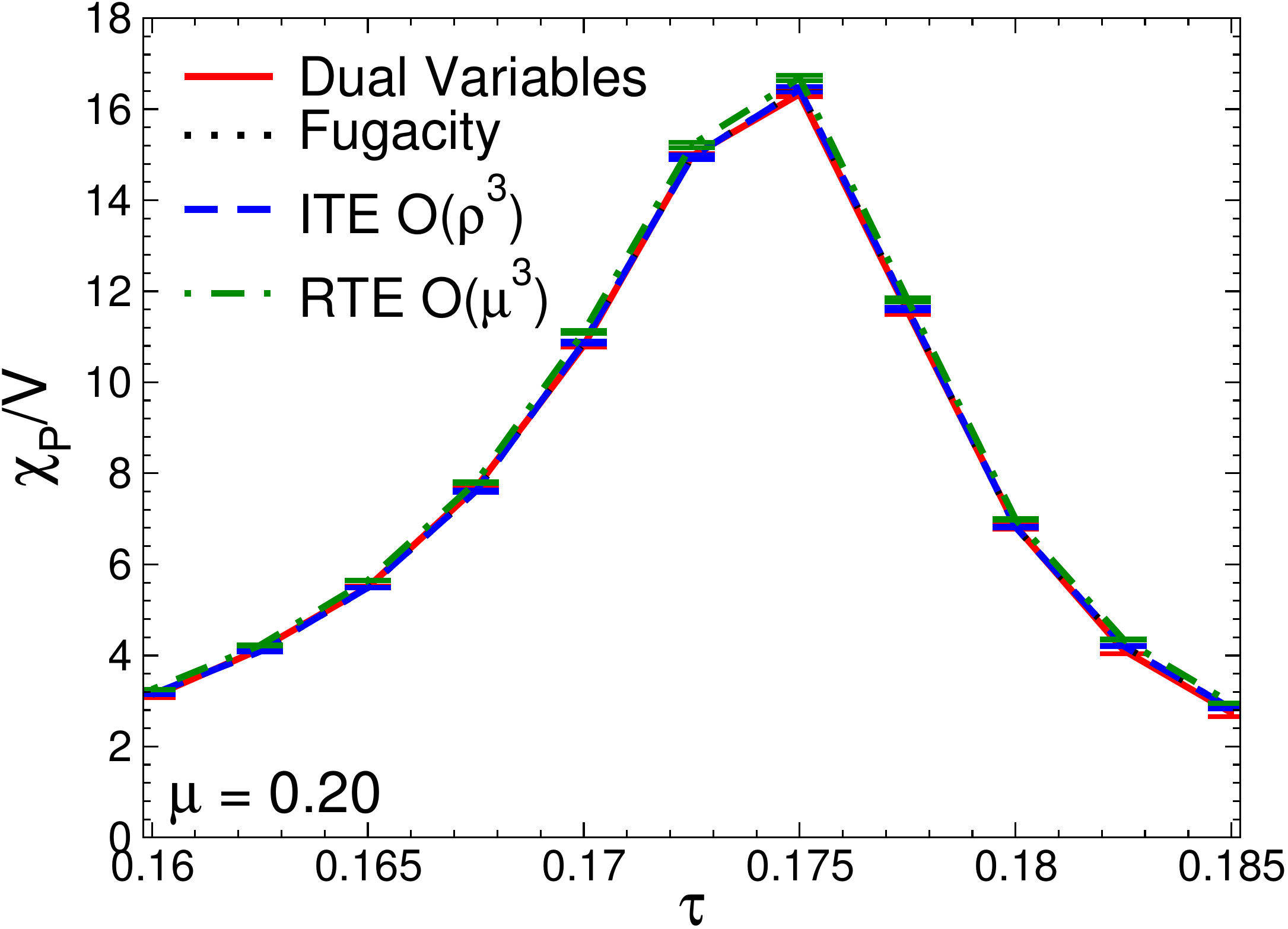}  
\hspace*{10mm}\includegraphics[height=38mm,width=50mm]{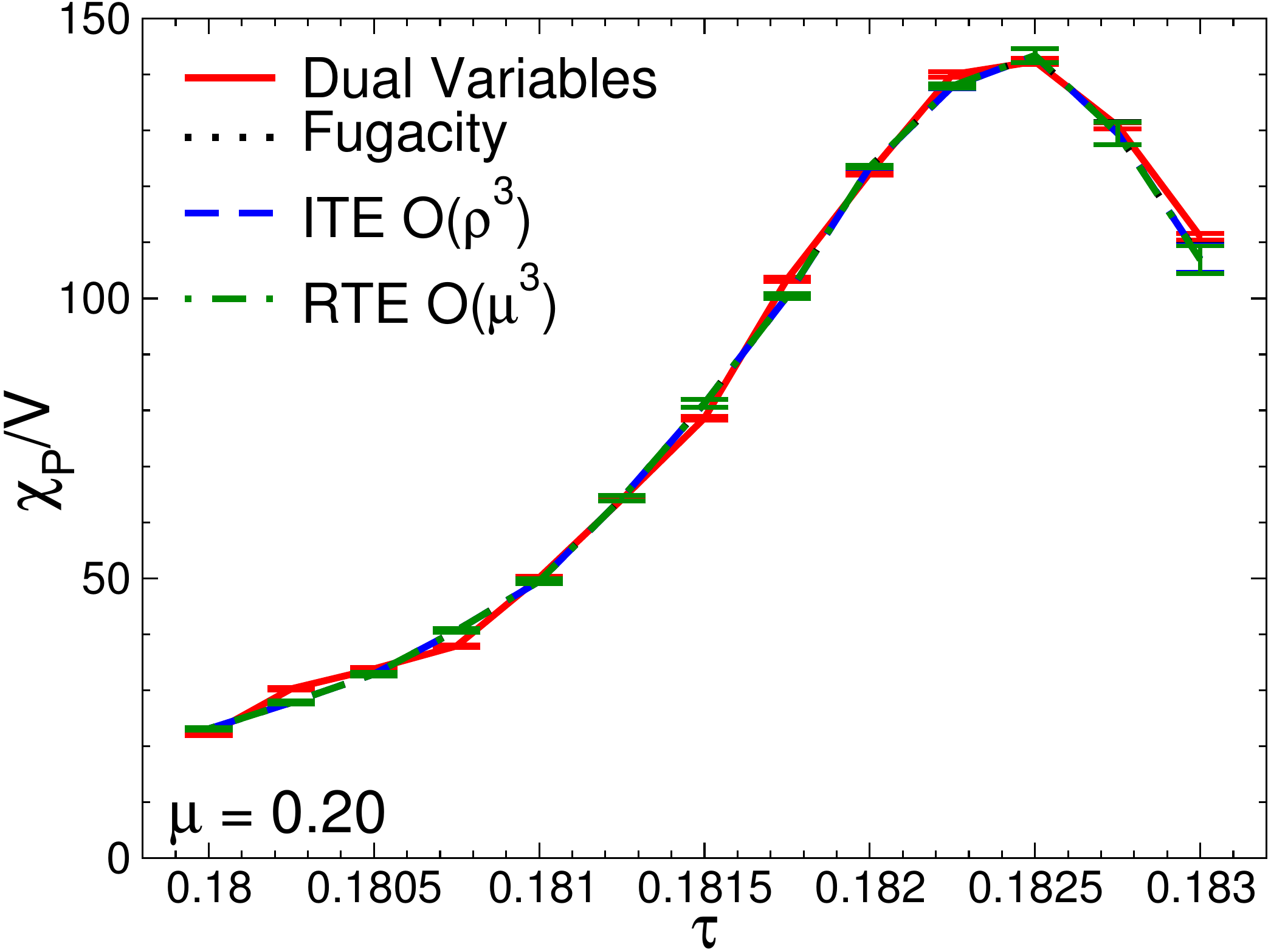}

\hspace*{8mm}\includegraphics[height=38mm,width=50mm]{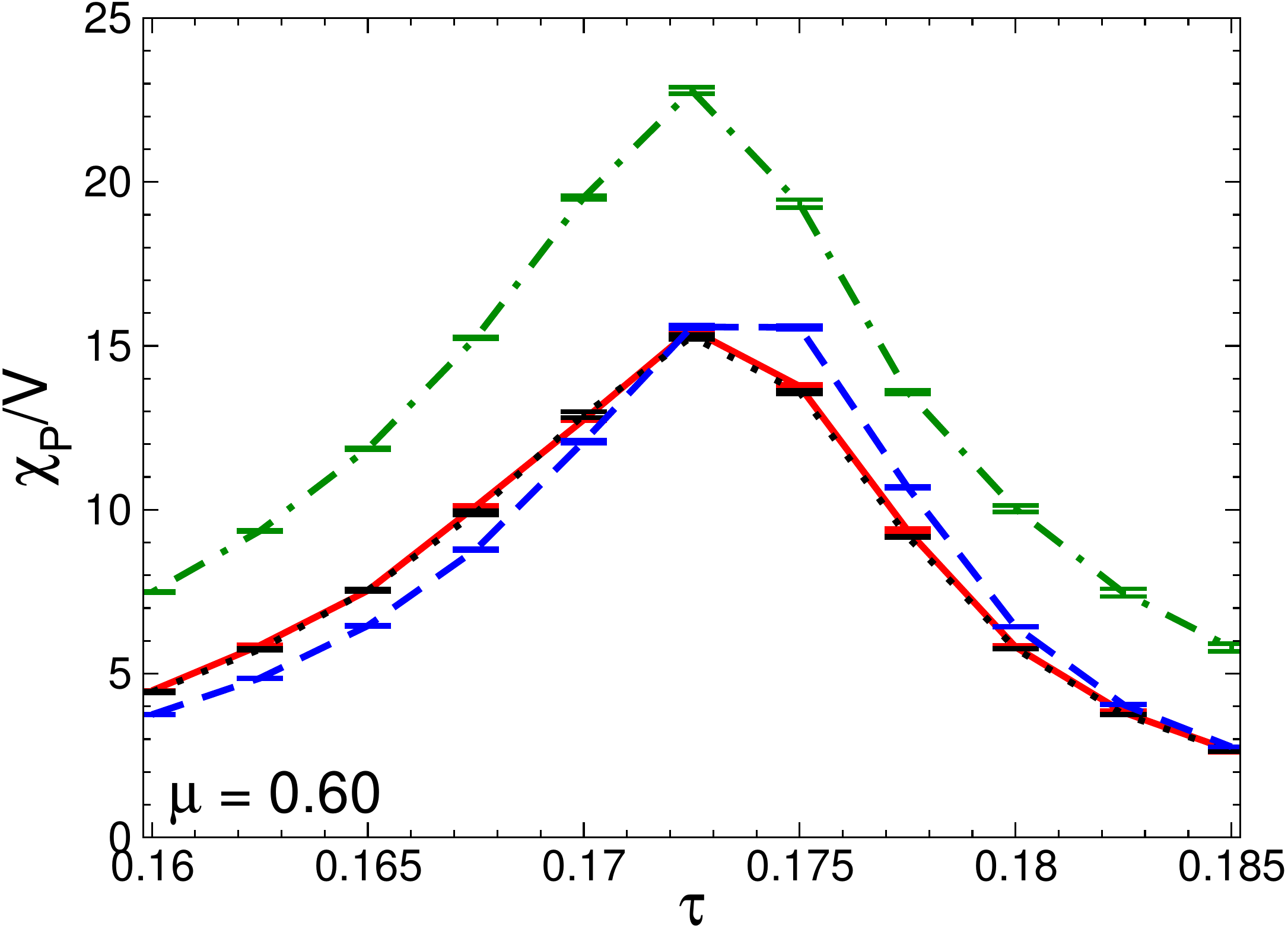}
\hspace*{9.8mm}\includegraphics[height=38mm,width=50.3mm]{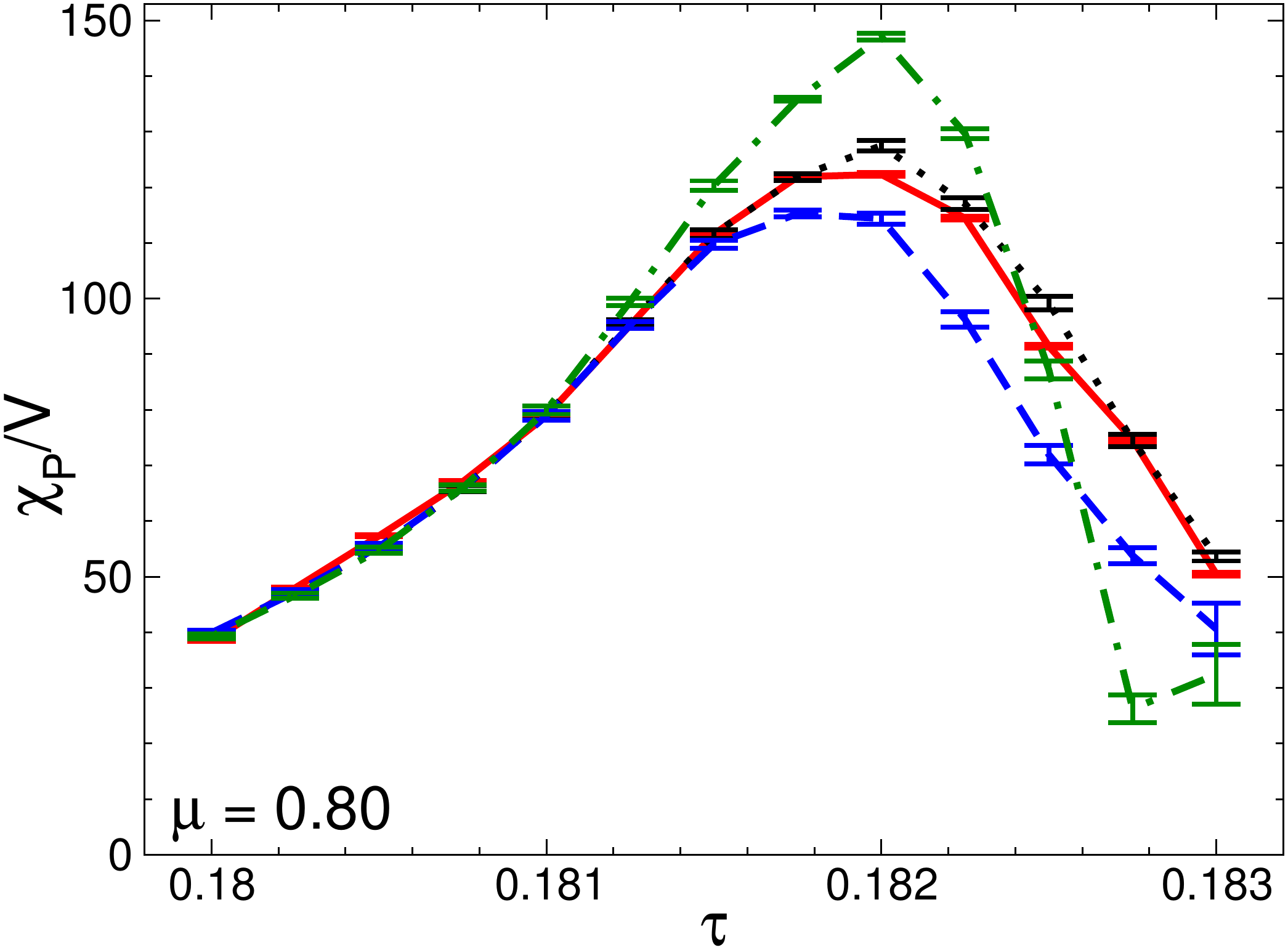}

\hspace*{8mm}\includegraphics[height=38mm,width=50mm]{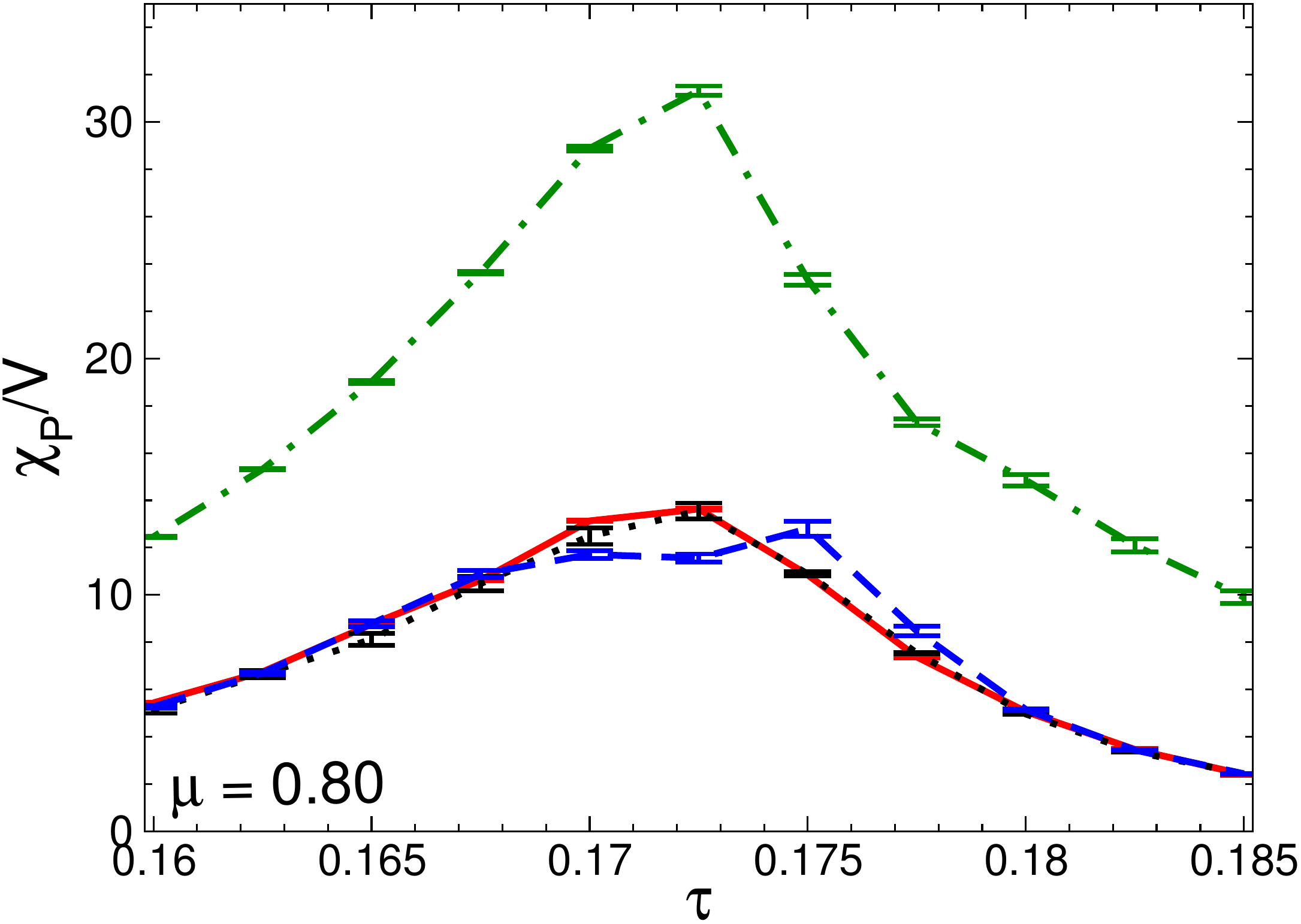}
\hspace*{9.8mm}\includegraphics[height=38mm,width=50.5mm]{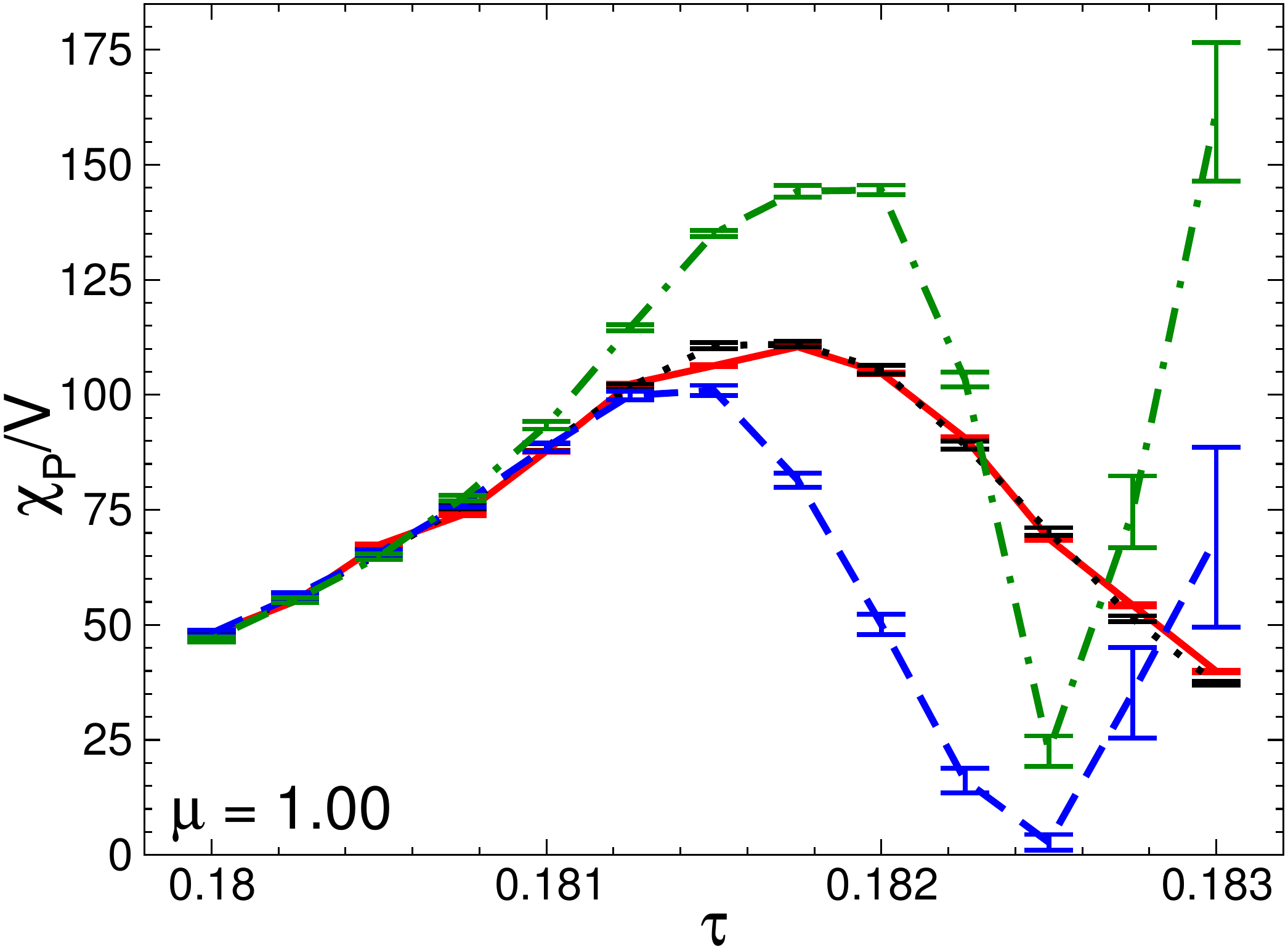}

\hspace*{8mm}\includegraphics[height=38mm,width=50mm]{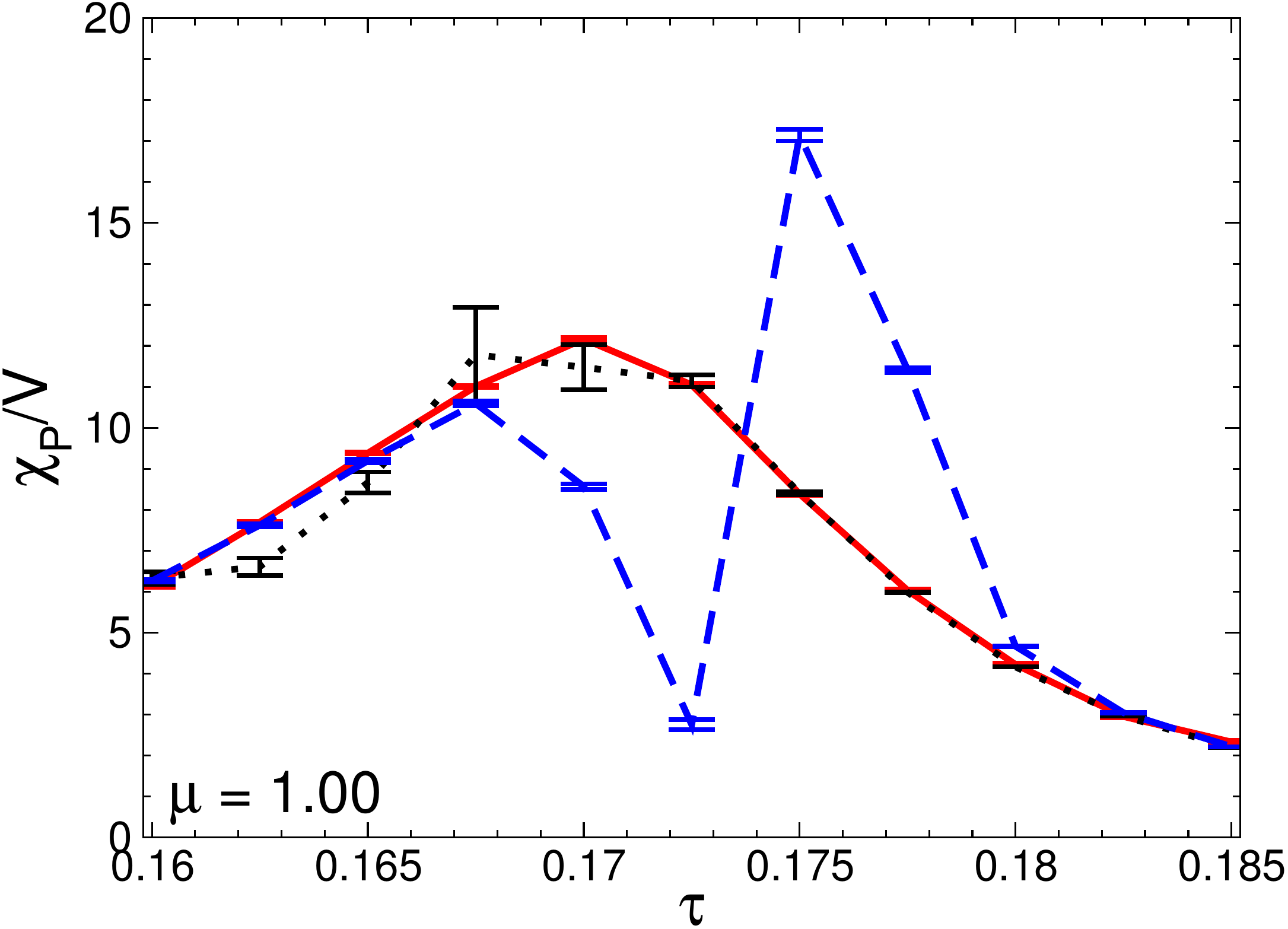}
\hspace*{9.5mm}\includegraphics[height=38mm,width=50.8mm]{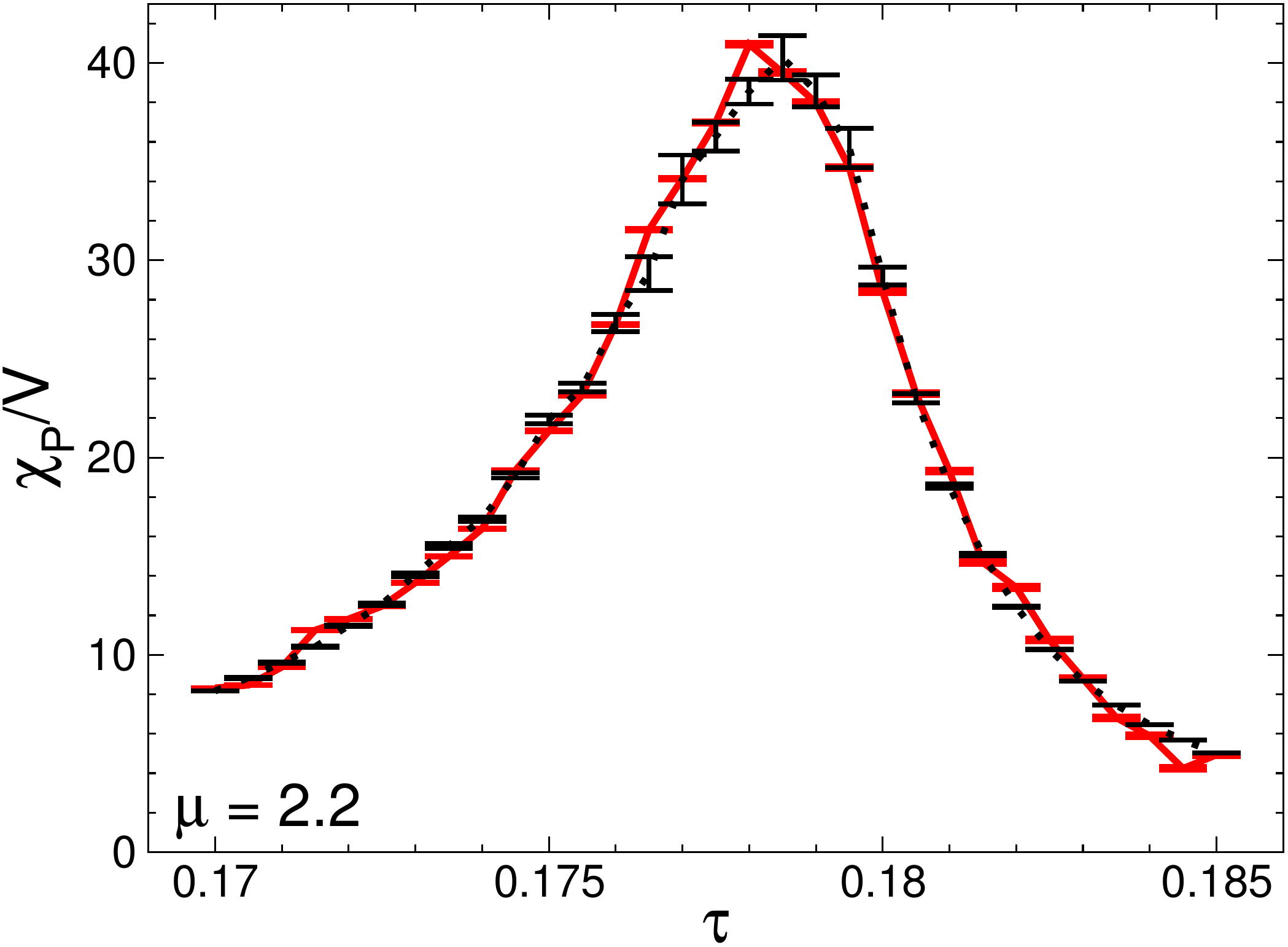}

\hspace*{7.4mm}\includegraphics[height=38mm,width=50.5mm]{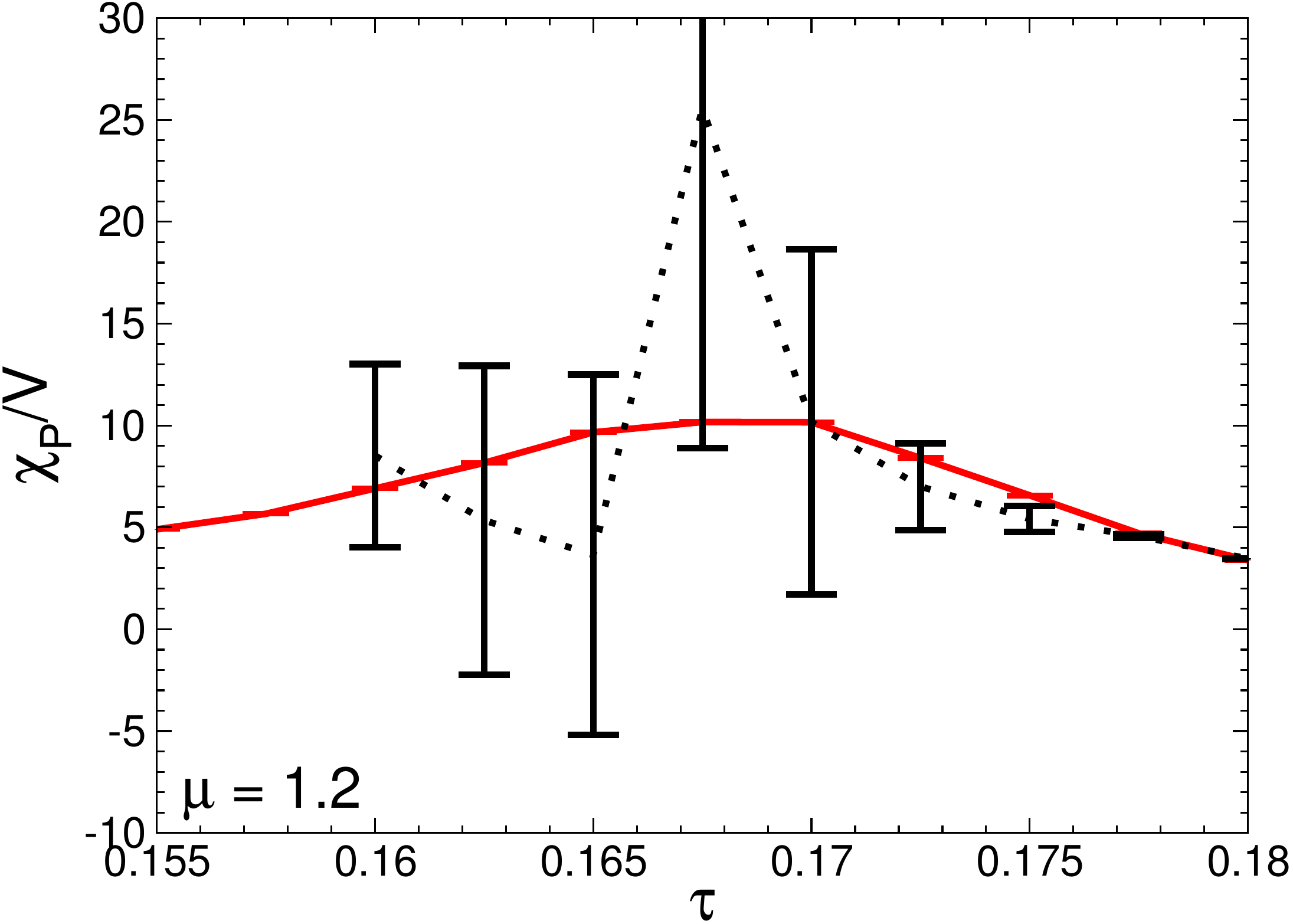}
\hspace*{9.8mm}\includegraphics[height=38mm,width=50.6mm]{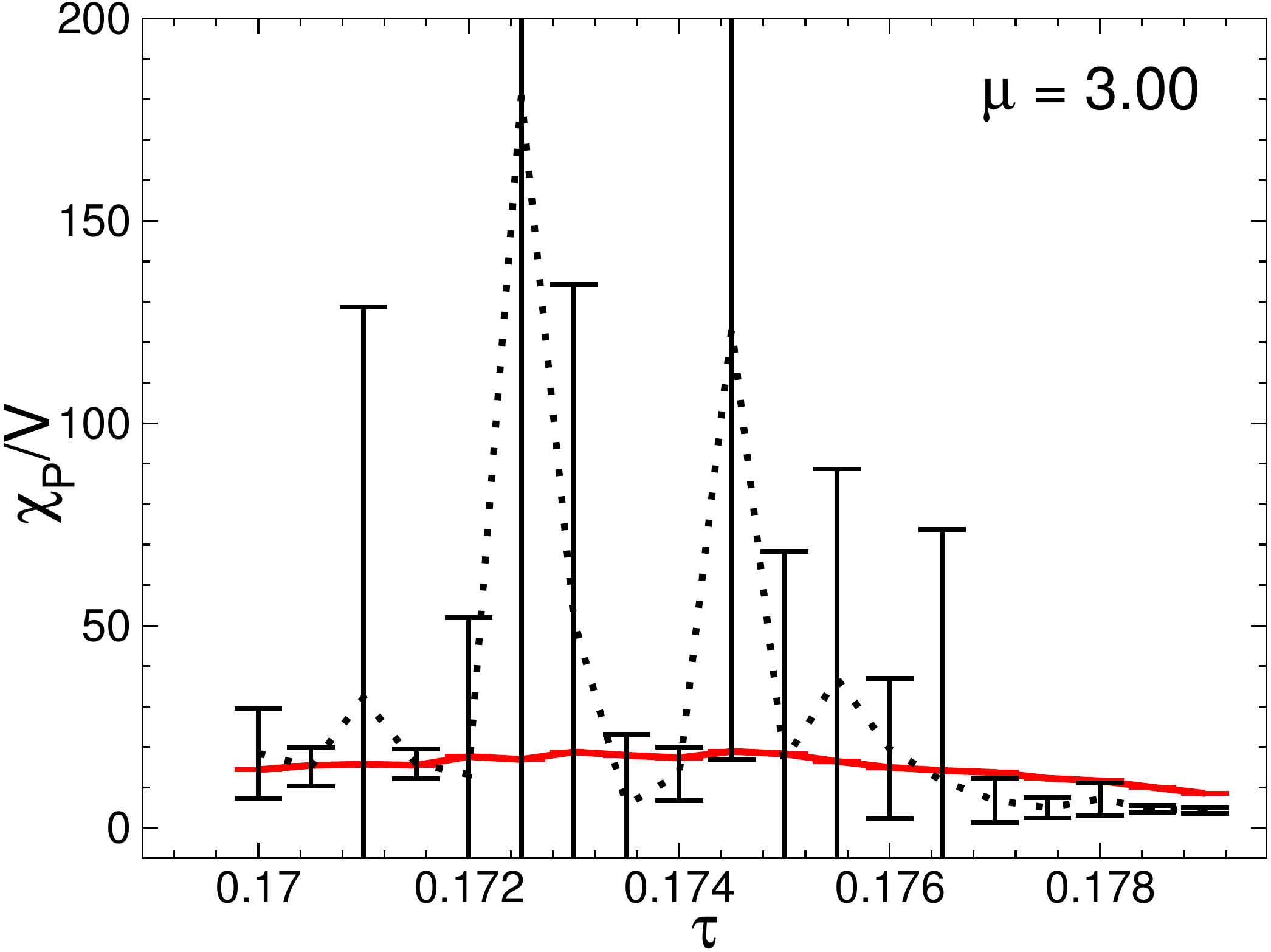}
 
\caption{Polyakov loop susceptibility $\chi_P$ at $ \kappa = 0.01$ (lhs.) and $ \kappa = 0.001$ (rhs.) 
from fugacity expansion, RTE, ITE and the dual simulation for different 
values of the chemical potential $\mu$.
\label{chiP_vergl}}
\end{figure}

For the two values of $\kappa = 0.01$ and $\kappa = 0.001$ and the corresponding values of $\mu$ listed above,
we show the results
for the particle number density $q/V$ (Fig.~\ref{q_vergl}), for the Polyakov loop expectation value $\langle P_x \rangle$ (labelled as
$P/V \equiv \langle P_x \rangle$ in Fig.~\ref{P_vergl}) and for the Polyakov loop susceptibility $\chi_P$ (Fig.~\ref{chiP_vergl}). 
All observables are plotted as a function of the temperature parameter $\tau$ and in the lhs.\ columns of Figs.\  
\ref{q_vergl}, \ref{P_vergl} and \ref{chiP_vergl} we show the $\kappa = 0.01$ results, while $\kappa = 0.001$ is shown on the rhs. 
In both columns the chemical potential increases from top to bottom with the respective values of 
$\mu$ indicated in the individual plots.

We begin the discussion of the results with the particle number density $q$ shown in Fig.~\ref{q_vergl} and the Polyakov loop 
$P$ (Fig.~\ref{P_vergl}), which are both first derivatives of $\ln Z$. For the $\kappa = 0.01$ data (lhs.\ columns in Figs.\ \ref{q_vergl} 
and \ref{P_vergl}) we find that all three series expansions are properly representing the dual simulation data up 
to $\mu = 0.6$. Above that the two Taylor expansions RTE and ITE start to deviate, while the fugacity expansion gives a proper 
representation up to $\mu = 1.0$. For the largest chemical potential we show, $\mu = 1.2$, all three series fail to reproduce the 
dual simulation data. For the $\kappa = 0.001$ data (rhs.\ columns in Figs.\ \ref{q_vergl} and \ref{P_vergl}) the situation is very
similar: The two Taylor expansions reproduce the dual simulation data only for small chemical potential, while the fugacity
expansion has a larger interval of good convergence, at least up to $\mu = 2.2$ for the $\kappa = 0.001$ data, and only for the largest 
value of $\mu$ we display, the failure of the fugacity series becomes manifest. 

It is interesting to note, that a comparison of the range of convergence of the fugacity series with Fig.~\ref{CAP} for the
severity of the complex action problem indicates, that only the fugacity series seems to be predominantly limited by  
the complex action 
problem. For both $\kappa = 0.01$ and $\kappa = 0.001$ we see that the fugacity series is reliable for the entire 
range of $\mu$ values where $\langle e^{i 2 \phi} \rangle_{p.q.} > 0.1$  (and for $\kappa = 0.01$ even slightly further). 
This is different for the two Taylor expansions, which break down already at values of $\mu$ smaller than what would look 
doable from the $\langle e^{i 2 \phi} \rangle_{p.q.}$ data in Fig.~\ref{CAP}. This underlines that the fugacity expansion has a larger
$\mu$-range of applicability than the Taylor series.

For the second derivative observable $\chi_P$ shown in Fig.~\ref{chiP_vergl}, we find essentially the same behavior as
for the first order derivatives $q$ and $P$, i.e., the two Taylor expansions break down considerably earlier than the fugacity
series, where the applicability of the latter seems to be limited mainly by the complex action problem, while the Taylor series
seem to have additional convergence issues, in particular fluctuations from higher order terms. 
However, a different feature is that for $\chi_P$ the improved Taylor series ITE
seems to be slightly better than the RTE, in particular at $\kappa = 0.01$.
 
\section{Summary}

In the work reported here, we assess fugacity-, regular Taylor- and improved Taylor expansion  
in the $\mathds{Z}_3$ spin model and compare the results from the series expansions in the chemical potential $\mu$
to the outcome of simulations in the dual formulation where the complex action problem is absent. 
We study the reliability of the series expansions in reproducing bulk observables at various values of the chemical potential. 

Our analysis shows that the fugacity expansion clearly outperforms the two Taylor series expansions for all parameter
values we studied. The convergence of the results seems to be limited exclusively by the severity of the sign problem,
while for the two Taylor series other issues, such as fluctuating terms from higher contributions limit the applicability. For some
parameter values we observed a slight superiority of the improved Taylor expansion ITE.

One goal of the current paper was to analyze in a QCD-related model system whether the much larger numerical cost of 
an implementation of the fugacity expansion is worth the effort when compared to Taylor expansion techniques. The results
presented here indeed suggest that the fugacity series could be a superior expansion also in QCD.

\vskip5mm
\noindent
{\bf Acknowledgments:} 
This work was supported by the Austrian Science Fund, 
FWF, DK {\sl Hadrons in Vacuum, Nuclei, and Stars} (FWF DK W1203-N16). Y.~Delgado is supported by
the Research Executive Agency (REA) of the European Union 
under Grant Agreement number PITN-GA-2009-238353 (ITN STRONGnet) and by {\sl Hadron Physics 2}. 
Furthermore this work is partly supported by DFG TR55, ``{\sl Hadron Properties from Lattice QCD}'' 
and by the Austrian Science Fund FWF Grant.\ Nr.\ I 1452-N27.

\end{document}